\documentclass[aps,pra,showpacs,twoside,twocolumn,10pt]{revtex4-1}
\usepackage[colorlinks=true, citecolor=red, urlcolor=blue ]{hyperref}
\usepackage{epsfig,newlfont,amssymb,amsfonts,amsmath,bm,subfigure,palatino,mathtools,amsthm,braket,times,soul}
\usepackage{color}
\usepackage{multirow}
\usepackage{ulem}
\definecolor{indiagreen}{rgb}{0.07, 0.53, 0.03}

\begin{document}

\title{
 Significance of Fidelity Deviation in Continuous Variable  Teleportation
} 
	
	\author{Ayan Patra\(^1\), Rivu Gupta\(^1\), Saptarshi Roy\(^{1,2,3}\), and Aditi Sen(De)\(^{1}\)}
	
\affiliation{\(^1\)Harish-Chandra Research Institute,  A CI of Homi Bhabha National
Institute, Chhatnag Road, Jhunsi, Prayagraj - 211019, India}
\affiliation{\(^2\) Quantum Information and Computation Initiative, Department of Computer Science, The University of Hong Kong, Pokfulam Road, Hong Kong.}

	\begin{abstract}

Performance of quantum teleportation is typically measured by the average fidelity, an overlap between the input and output states.  Along with the first moment,  we introduce the second moment of fidelity in CV teleportation, i.e., the fidelity deviation as the figures of merit to assess the protocol's efficiency. We show that CV states, both Gaussian and non Gaussian, can be better characterized by considering both average fidelity and fidelity deviation, which is not possible with only average fidelity.
Moreover,  we shed light on the performance of the teleportation protocol in two different input scenarios - one is when input states are sampled from constrained uniform distribution while the other one is Gaussian suppression of the input states which again lead to a different classification of CV states according to their performance. The entire analysis is carried out in noiseless and noisy scenarios with noise being incorporated in the measurement and the shared channels. We also report that one type of noise can make the protocol robust against the other one which leads to a `constructive effect' and identify the noise models which are responsible for decreasing average fidelity and increment in fidelity deviation.

%identify a noise model which is more responsible t

%We then explore a detailed comparison among different class of resources in terms of both figures of merit. We show that the comparison among different resources using these figures of merit is not as trivial as considering only the one-shot fidelity. Finally we let noise comes into the play. Considering two different noises- noise in homodyne measurement and in quantum channel, we notice that one type of noise makes the protocol robust against the other which leads to a 'constructive effect'. Moreover we found that one type of noise is more responsible for degrading the average fidelity while the other affects fidelity deviation more dominantly.
	\end{abstract}
	
	\maketitle
	
	\section{Introduction}
	\label{sec:intro}
Quantum teleportation (QT), discovered in 1993 \cite{Bennett_PRL_1993} is unmistakably one of the remarkable pieces of sorcery that quantum mechanics makes possible. After its proposal, it has tasted unprecedented levels of success both theoretically  \cite{Horodecki_PRA_1999,Horodecki_PLA_1996,Verstraete_PRL_2003, Braunstein_PRA_1995, Rivu_PRA_2021,Horodecki_PRL_1999} and experimentally \cite{Bouwmeester_Nature_1997,Ursin_Nature_2004, Krauter_Nature_2013, Boschi_PRL_1998, Valivarthi_Nature_2016, Wang_Nature_2015, Lee_Science_2011, Zhang_Nature_2006, Jin_Nature_2010,Furusawa_Science_1998} in its mere two decade long existence. The latest feather in the cap of experimental QT is undoubtedly the satellite-based setups that give rise to a possibility of  realizing  quantum information transmission at intercontinental distances \cite{Xia_QST_2017,Takesue_Optica_2015,Jin_Nature_2010}.  Interestingly, this tremendous progress and success in this field do not limit the research directions, but on the contrary, widens it. In the theoretical frontier, in the last couple of years alone, several new and interesting facets have emerged in this field 
%field of quantum teleportation like
which include port based quantum teleportation \cite{Studzinski_Nature_2017, Jeong_PRA_2020, Christandl_CMP_2021, Quintino_Quantum_2021}, fidelity enhancement of noisy QT using quantum switch \cite{Zhang_JOB_2001, Koga_PRA_2018, Zanin_OptExp_2021, Caleffi_IEEE_2020},  teleportation involving multiple parties \cite{telecloning, ASD_multipartite_channelcapacity_2010, telecloningCV, Saptarshihighway}, multiround quantum teleportation using weak measurements \cite{Saptarshi_arxiv_2019}, fidelity deviation in QT \cite{Bang_JPA_2018}.

Among these various avenues, let us briefly discuss  and elaborate on the importance of the idea of fidelity deviation in QT. Typically, the performance of QT is measured by the average fidelity. However, such a mean-based characterization has some limitations since  it cannot capture the fluctuations in fidelity with the various choices of inputs from the ensemble of states that are supplied for teleportation. For example, fluctuations become very important in situations where teleportation is used as an intermediate step in a quantum information processing task involving quantum gates. Since the performance of quantum gates  depends on the fluctuations of its input (that reaches the gate via QT) \cite{Jeong_PRA_2002,Raplh_PRA_2003}, the fidelity deviation must be taken into account on top of average fidelity for characterizing the quality of QT.  Noting its importance, several works have been carried out in investigating the role of fidelity deviation in QT \cite{Arkaprabha_JPA_2020,Arkaprabha_PRA_2020,Saptarshi_PRA_2020,Saptarshi_PRA_2022}. 

Continuous variable (CV) systems offer some distinct advantages over their discrete counterparts whereby they can overcome certain difficulties, like Bell-basis indistinguishability via linear optics \cite{pirandola_multiparty-tele_2015}. Furthermore, they can be prepared
with near perfect efficiency by using nonlinear interaction of a crystal with laser, and the only  imperfection can arise due to the varying intensity of laser light, resulting in a low squeezing parameter \cite{Andersen_PS_2016}, thereby making them potential systems
%. This makes CV systems one of the potential candidates for 
for implementing quantum information processing tasks.  Among the set of CV systems, Gaussian states hold a privileged position owing to their mathematical simplicity and experimental realizability \cite{Nokkala_Nature_2021,ferraro_arXiv_2005,Adesso_OSID_2014,Serafini_2017}.

Notably, it was in the Gaussian domain that the idea of CV teleportation was first conceptualized by Vaidman, Braunstein and Kimble (referred to as the VBK protocol) \cite{Vaidman94, Braunstein_PRL_1998}. 
%a good seven years after the idea of QT in discrete variable  was first introduced \cite{Bennett_PRL_1993}.  
From its inception, several directions have been explored in CV QT by varying the one-shot fidelity \cite{DellAnno_EPJS_2008,DellAnno_PRA_2007} and the average fidelity  \cite{Villasenor_arxiv_2021}.  It includes the extension of the protocol to non-Gaussian regimes, exhibiting  that photon subtracted (PS) states can  outperform the two mode squeezed vacuum (TMSV) state according to the average fidelity  \cite{VILLASENOR_IEEE_2021,Wang_PRA_2015,Kitagawa_PRA_2006,Olivares_PRA_2003,Cochrane_PRA_2002,Opatrny_PRA_2000,DellAnno_EPJS_2008}, incorporating noise \cite{DellAnno_PRA_2007,DellAnno_PRA_2010,Villasenor_arxiv_2021,Johnson_PRA_2002}, constructing CV QT networks \cite{Yonezawa_Nature_2007,Furusawa_IEEE_2005,vanLoock_PRL_2000}, understanding the relationship between measures of quantum correlations and the fidelity  \cite{Himadri_JPB_2015,Kim_PRA_2013,Nha_PRL_2012,Menzel_PRL_2012,adesso_arXiv_2007,Takei_PRA_2006,Doli_PRA_2003,Adesso_PRL_2007} (a problem which is considerably well understood in the discrete case \cite{Horodecki_PLA_1996,Horodecki_PRA_1999}) and many more. To show quantum advantage, the classical threshold for the coherent state is shown to be at most half
%$\mathcal{F} \leq 0.5$ 
and quantum resources are known to beat the optimum measure-prepare strategy for moderate to high values of the squeezing parameter (see Fig. \ref{fig:noiseless_avfid_r}). For a more detailed review of the literature, see  \cite{Pirandola_LP_2006}. 

%Till now, research on the teleportation fidelity has mainly been concentrated on the variation of the one-shot fidelity \cite{DellAnno_EPJS_2008,DellAnno_PRA_2007} and the average fidelity  \cite{Villasenor_arxiv_2021} with the resource squeezing. The resources considered have been both Gaussian and non-Gaussian. It is well known that the average fidelity increases with an increase in the resource squeezing \cite{DellAnno_PRA_2010_2,DellAnno_PRA_2010}
%. It has also been established that the
%while non-Gaussian . \\

In this work, we focus on two independent aspects of CV quantum teleportation. On one hand,  our work focuses on assessing the quality of the protocol with respect to the variation in input energy. Specifically, we consider input states coming from different energy distributions - uniform distribution having a finite threshold in the maximum permissible energy to avoid divergence and Gaussian distributions characterized by a specific variance. For example, we study how the average fidelity scales with different input distributions and examine the regimes at which quantum advantage is apparent, since the classical bound also varies for different energy constraints.  Notice that in a recent work \cite{Wilde_2022}, optimal input states are determined by considering energy-constrained fidelity between ideal scenarios and additive noise induced channels, which can be the case in experiments. From a different perspective, some of us have recently shown that if the input states are derived from non-uniform distributions, instead of being distributed over the entire Bloch sphere, the teleportation protocol in terms of its average fidelity and fidelity deviation can be improved \cite{Saptarshi_PRA_2022}. Motivated by these results, in this work, we consider the uniform distribution of input states which are constrained in energy and show that distributions with lower energy cut-offs can aid in the teleportation protocol since less resource squeezing is required for optimal transfer of such states. In this new paradigm, we also compute the entanglement-free (measure-prepare) bound on QT to show where quantum advantage is manifested.

On the other hand, we introduce the concept of the second moment of the fidelity statistics, the fidelity deviation, in CV systems which quantifies how well a given resource aids in the teleportation of different states coming from a given ensemble. A lower value of the fidelity deviation indicates that the resource is capable of transferring various input states with fidelities very close to the average fidelity. This is essential, since even if the average fidelity is high, a large deviation means that some states might still be teleported with suboptimal fidelity. 

Our aim here is to determine the performance of QT by examining both the average fidelity and fidelity deviation and classify the CV resource states, both Gaussian TMSV as well as photon added and subtracted states according to their performances. Specifically, we report that  contrary to the known results, TMSV states turn out to be better suited for CV QT than the PS states in many situations in the presence or absence of noise.
The investigations are carried out for different input states, coherent, squeezed, and squeezed coherent states when  entangled channels are shared.
In a noisy regime,  we show that  noise in measurement can be circumvented by a moderate amount of noise in channels, which we refer to as a constructive effect irrespective of the input energy. Moreover,  we observe that  noise in measurement at the sender's end has adverse effects on the quality of CV QT in terms of the average fidelity and its deviation compared to the noise in the channels. However, both the noise  models have more detrimental effects on non-Gaussian states than the Gaussian ones, thereby establishing the TMSV state as a  suitable channel for teleportation in presence of high input energy.

If we now compare the results obtained in CV systems with the known results for two qubit systems \cite{Arkaprabha_JPA_2020,Arkaprabha_PRA_2020,Saptarshi_PRA_2020,Saptarshi_PRA_2022},
the principal point of difference 
%of the work presented in this manuscript with the results obtained in \cite{Arkaprabha_JPA_2020,Arkaprabha_PRA_2020,Saptarshi_PRA_2020,Saptarshi_PRA_2022} 
is the dimension of the systems under consideration. %that all  these studies are limited to discrete variable systems, particularly dealing with qubits. 
Incorporating the idea of fidelity deviation in the continuous variable  paradigm  is qualitatively different from teleportation with qubits. Hence although the known works provide a wider perspective for qubit teleportation,  this knowledge does not provide substantial intuition in the CV case. 
In Ref. \cite{Arkaprabha_JPA_2020}, fidelity deviation was studied for generic two qubit states and the existence of resources for which the fidelity deviation could have vanishing values are identified, thereby making them universal for the teleportation protocol. In our work, we study the properties of fidelity deviation for two mode entangled resource states and arrive at a hierarchy between different Gaussian and non-Gaussian states in terms of their fidelity deviation when the input energy is fixed within a given range. It is, however, an open question as to whether universal continuous variable resource states exist, which can teleport states without any fluctuations in the average fidelity.
It is also interesting to characterize states in terms of their fidelity deviation based on state properties such as linear entropy, concurrence etc and show that such properties may also dictate the teleportation protocol  \cite{Arkaprabha_PRA_2020}. Instead of establishing such a connection, we mainly focus here on the interplay between the average fidelity and the fidelity deviation of states for inputs belonging to an ensemble of fixed energy. We believe that our work helps to identify resources suitable for teleportation depending on the energy of the state to be teleported even in the presence of noise.
Recently, combining the two moments of fidelity, a new performance measure called teleportation score was introduced  \cite{Saptarshi_PRA_2020}, for two qubit systems. Such a measure for CV teleportation is too intricate since it depends on several state properties. In our work, we leave it as an open question whether such a measure exists for CV systems as well.

%noise models which leads to a sharp decrements of the average fidelity and increments of its deviation. 

%and determine the best possible resource suitable for the protocol.
%The analysis is carried out both in the noiselss and noisy scenarios. We will show that fidelity deviation along with average fidelity can unfold certain issues which cannot be addressed by considering fidelity alone. 

%\textcolor{red}{eikhan ta aro likhte hobe. }

% we determine the fidelitites of teleporting unknown squeezed, coherent and squeezed-coherent states with the aid of an entangled resource (Gaussian and non-Gaussian) and compare the same with the classical bound which is calculated by setting the squeezing of the resource state to zero. We find that the classical threshold depends on the energy cutoff of the uniform distribution and also on the standard deviation of the Gaussian distribution.

%In this work we explore the role of fidelity deviation in  CV QT. We perform our investigation for both Gaussian and non-Gaussian resources in the presence and absence of  noise. We propose a practical way of constructing the average fidelity and deviation using finite energy constraints....

The paper is organized as follows. Before presenting the results,   we introduce the monitors which can assess the performance of QT,  the role of input energy in the performance and describe briefly the classical limit in each situation (Sec. \ref{sec:fig_merit}). The trends of average fidelity and fidelity deviation for noiseless CV teleportation with respect to different inputs and resource states are presented in Sec. \ref{sec:noiseless}. The hierarchies among states according to the fidelity deviation are studied in Sec. \ref{sec:deviation} for the noiseless case. In Sec. \ref{sec:noisy}, we investigate the effects of noise on the performance of CV QT by considering the average fidelity while the behavior of fidelity deviation in presence of noise 
is discussed in Sec. \ref{sec:fid_dev_noisy}. 
Finally, we make the concluding remarks  in Sec.  \ref{sec:conclu}.

\section{Regularized Figures of merit}
\label{sec:fig_merit}

%Define various elements of CV quantum teleportation. Figures of merit e fidelity deviation er defn.

Continuous variable  systems are characterized by canonically conjugate observables, say $X$ and $P$, possessing a continuous spectrum. The system Hamiltonian for $N$ such pairs, each of which corresponds to a different mode, reads as
\begin{eqnarray}
H = \frac{1}{2}\sum_{k=1}^N (X_k^2 + P_k^2) = \sum_{k=1}^N a_k^{\dagger} a_k + \frac{N}{2},
\end{eqnarray}
where $k$ denotes the mode, while $a_k$ and $a_k^\dagger$ represent the photon annihilation and creation operator respectively with 
\begin{eqnarray}
a_k =  \frac{X_k + i P_k}{\sqrt{2}}, ~\text{and}    ~a_k^\dagger =  \frac{X_k - i P_k}{\sqrt{2}},
\end{eqnarray}
where $i = \sqrt{-1}$.
When a single mode state,  $|\psi_{\mathrm{in}}\rangle$, has to be teleported through a CV channel, the overlap between the output state after implementing the protocol, $\rho_{\mathrm{out}}$, and the input state $|\psi_{\mathrm{in}}\rangle$, referred to as the fidelity $f(|\psi_{\mathrm{in}}\rangle) = \langle\psi_{\mathrm{in}}|\rho_{\mathrm{out}}|\psi_{\mathrm{in}}\rangle$ measures the efficacy of the protocol. When the standard CV teleportation scheme is followed \cite{Braunstein_PRL_1998}, the fidelity can be expressed as \cite{chizhov_2002}
\begin{eqnarray}
f_{|\psi_{\mathrm{in}}\rangle} = \frac{1}{\pi}\int d^2\alpha \chi_{in}(-\alpha)\chi_{\mathrm{out}} (\alpha),
\end{eqnarray}
where $\chi_{\mathrm{out}}(\alpha)=\chi_{\mathrm{in}}(\alpha)\chi_{\mathrm{res}}(\alpha^\ast,\alpha)$ \cite{Marian_2006} with $\chi_\rho(\alpha) = \mbox{tr}(\rho D(\alpha))$, $D(\alpha)$ being the displacement operator, and $\chi_\rho(\alpha)$ is the characteristic function of the single-mode state $\rho$.

 In our work, we primarily choose $|\psi_{\mathrm{in}}\rangle$ to be a single mode pure Gaussian state \cite{Chiribella_PRL_2014}. Recall that the most general single mode pure Gaussian states are the squeezed-coherent states \cite{ferraro_arXiv_2005}, and therefore, most generally,
\begin{eqnarray}
|\psi_{\mathrm{in}}\rangle \in \{S(\xi)D(\beta)|0\rangle\} ~\forall \xi,\beta \in \mathbb{C},
\end{eqnarray}
where $S(\xi) = \exp{\big[\frac{1}{2}(\xi^\ast \hat{a}^2 - \xi \hat{a}^{\dagger2})\big]}$ represents the single mode squeezing operator with $\xi = \epsilon~e^{i\theta}$ in which $|\xi| = \epsilon$
 represents the squeezing strength while $\theta$ denotes the squeezing angle.  Here  $\beta=be^{i\phi}$ is the displacement parameter. Choosing $|\psi_{\mathrm{in}}\rangle$ uniformly from the above ensemble is unphysical since it leads to divergent energies, which can be noted from the average energy of $|\psi_{\mathrm{in}}\rangle$, 
 \begin{eqnarray}
 E_{|\psi_{\mathrm{in}}\rangle} = \langle \psi_{\mathrm{in}}|H|\psi_{\mathrm{in}}\rangle = b^2 + \sinh^2 \epsilon.
 \label{eq:avgen}
 \end{eqnarray}
 This divergence can be prevented by imposing a  distribution $p(\beta, \xi)$ with $\frac{1}{\mathcal{N}}\int p(\beta, \xi) d^2\beta ~d^2\xi = 1$ on the choice of $|\psi_{\mathrm{in}}\rangle$ such that the average energy for the distribution of input states
 \begin{eqnarray}
 E_{avg} = \frac{1}{\mathcal{N}} \int p(\beta,\xi) E_{|\psi_{\mathrm{in}}\rangle} d^2\beta ~d^2\xi < \infty. 
 \label{eq:avgen_dist}
 \end{eqnarray}
Endowed with this prescription for taming the divergences, we  classify the performance of CV quantum teleportation using the first two moments of fidelity, referred to as the average fidelity \cite{DellAnno_PRA_2010}, given by  
%Borrowing inspiration from the discrete variable case, the average fidelity $F$ can be formally written down as
\begin{eqnarray}
\mathcal{F}  = \frac{1}{\mathcal{N}}\int p(\beta,\xi) f_{|\psi_{\mathrm{in}}\rangle} d|\psi_{\mathrm{in}}\rangle, 
\label{eq:av_fid}
\end{eqnarray}
with $\mathcal{N} = \int p(\beta,\xi) d|\psi_{\mathrm{in}}\rangle$ while 
the corresponding fidelity deviation \cite{Bang_JPA_2018} reads as
\begin{eqnarray}
\Delta \mathcal{F} = \sqrt{\langle f_{|\psi_{\mathrm{in}}\rangle}^2 \rangle - \mathcal{F}^2},
\label{eq:fid_dev}
\end{eqnarray}
where $\langle f_{|\psi_{\mathrm{in}}\rangle}^2 \rangle = \frac{1}{\mathcal{N}}\int p(\beta, \xi) f_{|\psi_{\mathrm{in}}\rangle}^2 d|\psi_{\mathrm{in}}\rangle$.
Notice that in the discrete case, the measure ``$d|\psi_{\mathrm{in}}\rangle"$ implies the entire space of inputs chosen uniformly from the Hilbert space of the relevant dimension, i.e., $p(\beta, \xi)$ is an uniform distribution.
%However, in the CV case, uniformly choosing all inputs implies that the average energy of the input ensemble diverges. 
Here in CV systems,  we choose the measure $d|\psi_{\mathrm{in}}\rangle$ with reasonable cut-offs as mentioned before, making the average energy of the input ensemble  finite. This allows us to construct regularized versions of average fidelity and fidelity deviation that are free from typical divergences arising due to infinite dimensional systems.

In our analysis,  we consider two different realizations of $p(\beta,\xi)$, one with a finite cut-off in energy which we call the constrained uniform distribution,
%while the other adopts the scheme of 
and the other with a Gaussian suppression, respectively given by
 \begin{eqnarray}
&& p_C(\beta,\xi) = \left\{
 \begin{array}{cc}
 \mbox{constant}, &   E_{|\psi_{\mathrm{in}}\rangle} \leq \mathcal{E}\\
 0, & E_{|\psi_{\mathrm{in}}\rangle} > \mathcal{E}
\end{array}\right.,
\label{eq:p_uni}\\
&& p_G(\beta,\xi) =  e^{-\frac{b^2}{\sigma_c}}e^{-\frac{\epsilon^2}{\sigma_s}}.
\label{eq:p_gau}
 \end{eqnarray}
 Both of these distributions rectify the divergent issues. 
From Eq. \eqref{eq:avgen}, the condition in Eq. \eqref{eq:p_uni} can be rewritten as $b^2 + \sinh^2 \epsilon \leq \mathcal{E}$. In this case, the average fidelity can be modified as
%may be computed as
\begin{eqnarray}
\mathcal{F} = \frac{1}{\mathcal{N}} \int_{\epsilon = 0}^L && \int_{\theta = 0}^{2\pi} \int_{b = 0}^{\sqrt{\sinh^2L - \sinh^2\epsilon}} \int_{\phi = 0}^{2\pi}  f_{|\psi_{\mathrm{in}}\rangle} ~d|\psi_{\mathrm{in}}\rangle, \nonumber \\
\label{eq:av_fid_sq_coh}
\end{eqnarray}
where $d|\psi_{\mathrm{in}}\rangle = b~\epsilon~d\epsilon~d\theta~db~d\phi$ and the integral over the displacement parameter $b$ runs from zero to the part of the total energy not carried by the squeezing. We assume that the total energy is given by $\sinh^2 L = \mathcal{E}$, where $L$ is the maximum value that the squeezing parameter of the state, $\epsilon$, can possess. The integration measure in Eq. \eqref{eq:av_fid_sq_coh}  becomes $d\ket{\psi_{\mathrm{in}}}=d^2\xi = 2 \pi \epsilon d\epsilon$ for the squeezed state, whereas for the coherent state, it is $d\ket{\psi_{\mathrm{in}}}=d^2\beta = b~ db~ d\phi$. Since it is hard by state of the art experiments to achieve squeezing beyond $r = 1.6$ \cite{Vahlbruch_PRL_2016}, we accordingly fix the energy threshold for the squeezed state as $L_\xi = 1.6$ 
%\textcolor{red}{we shud refer some paper} 
such that $\sinh^2\epsilon \leq \sinh^2L_\xi$. In order to facilitate comparison, we also consider the energy cut-off for the coherent state to be the same, due to which $|L_\beta|^2 = \sinh^2L_\xi$ even though technically, it can possess relatively high energy.

For the Gaussian distribution of squeezed-coherent states, the integrals for computing average fidelity and fidelity deviation get  simplified forms since $p_G(\beta,\xi)$ acts independently on the coherent and squeezed sectors owing to its product structure. Note that such simplification is not possible with uniform distribution having energy thresholds. Using Eq. (\ref{eq:av_fid}) with the condition in Eq. (\ref{eq:p_gau}), the average fidelity for teleporting squeezed-coherent states can be computed as 
\begin{eqnarray}
    \mathcal{F} = \frac{1}{\mathcal{N}} \int_{\epsilon = 0}^\infty\int_{\theta = 0}^{2\pi} \int_{b = 0}^\infty \int_{\phi = 0}^{2\pi}  \nonumber\\
 f_{|\psi_{\mathrm{in}}\rangle} \exp(-\frac{\epsilon^2}{\sigma_s})\exp(-\frac{b^2}{\sigma_c}) ~d|\psi_{\mathrm{in}}\rangle,
    \label{eq:av_fid_sqzcoh_gaussian_dist}
\end{eqnarray}
where $\mathcal{N}=(\pi\sigma_s)(\pi\sigma_c)$, with $\sigma_s$ and $\sigma_c$ being the variance corresponding to the input squeezing  and the displacement parameters respectively. Throughout the paper, we have considered variance to be the Gaussian distribution parameter and both the figures of merit are analyzed with respect to variance in case of sampling of input states from Gaussian distribution.\\

Using Eq. (\ref{eq:avgen_dist}), the average input energy is computed to be $(\sigma_c + \frac{1}{2}e^{\sigma_s}\sqrt{\pi\sigma_s}\text{Erf}(\sqrt{\sigma_s}))$, where the first term represents the average energy for input coherent state and the second term corresponds to the average energy for input squeezed state individually. Here Erf is the error function given by $\text{Erf}(x)=\frac{2}{\sqrt{\pi}}\int_0^x{e^{-t^2}}~dt$. Since  the average energy depends both on $\sigma_s$ and $\sigma_c$ and increases with them,   we take the range of $\sigma_s$ and $\sigma_c$ up to $5.0$ and $10.0$ respectively in order to capture all possible prime features that the figures of merit can exhibit, with respect to the average input energy. Now we briefly discuss how these reasonable cutoffs of the average input energy relate to the practical teleportation experiment.

The energy constraint considered in the manuscript does indeed hold true for practical teleportation experiments. In any realistic CV teleportation setup, the input states have some cut-off in their energy because states with a very high energy are unphysical and are typically difficult to prepare in experiments. Such issues have also been addressed in other works of CV quantum teleportation in literature  \cite{VILLASENOR_IEEE_2021,Wang_PRA_2015,Kitagawa_PRA_2006,Olivares_PRA_2003,Cochrane_PRA_2002,Opatrny_PRA_2000,DellAnno_EPJS_2008} where only the average fidelity is computed. We have generalized CV teleportation via computing the first two moments of fidelity (the average fidelity and the fidelity deviation) under such constraints. Summarizing, we consider two instances of such energy-based constraint:
\begin{enumerate}
    \item The first is constrained uniform distribution, where the input set is constituted only by states less than a particular amount of energy. Inside this energy constrained set, the states are chosen uniformly. We motivate such a choice from qubit teleportation, where the input states for teleportation are considered to come from an unconstrained uniform distribution.
    \item The second is Gaussian distribution of the energy of the inputs. It has already been considered in other works of CV teleportation  involving only the average fidelity.  
\end{enumerate}
In a practical situation, where teleportation is used as an intermediate in a quantum circuit involving CV states, the states to be teleported are either prepared or generated by some prior processes. In both cases, the ability of the source to produce a very high energy states is restricted. For example, if one considers CV entanglement swapping \cite{Zukowski_PRL_1993, Bose_PRA_1999} (which is essentially teleporting a part of the entangled state \cite{Zukowski_PRL_1993, Torres_PRA_2014, Jin_Nature_2015}), the states which are shared among three different laboratories are a pair of two mode squeezed vacuum. Experimentally, the TMSV with highest possible entanglement that can be created so far constitutes an average energy per mode of $5.643$ in natural units \cite{Vahlbruch_PRL_2016}. These energy restrictions are ubiquitous in the CV regime, where, due to dimensional unboundedness, maximal values of extensive quantities diverge. Therefore, to make reasonable predictions, one must impose energy constraints.

Typically the input state at the sender’s side is completely unknown to the sender. However, it is reasonable to expect that the sender has some prior knowledge about the energy range of the input ensemble to be able to prepare the resource state and design a suitable strategy for optimal fidelity \cite{Vedral_PRA_2000,Saptarshi_PRA_2022}. Therefore, the sender does not require to perform any additional measurement on the input state to determine its energy and can simply implement the standard teleportation protocol once the state is supplied. The prior knowledge of the input energy ensemble thus allows the sender to teleport the input state in real CV experiments. One must keep in mind that the concept of prior knowledge in CV teleportation is fundamentally different from that in the discrete case. This is because the energy constraint which serves as prior knowledge is a physical requirement for the protocol to be implemented. It has to be present, contrary to the discrete case \cite{Saptarshi_PRA_2022} where prior knowledge is an additionally imposed assumption. In addition to this, one can also consider that the sender knows which kind of state has to be teleported i.e., whether the state is from an ensemble of squeezed states or coherent states or squeezed coherent states. This would allow the sender to further tailor the resource for successful teleportation. Note that this assumption is a standard practice in the literature of CV teleportation \cite{Vaidman94, Braunstein_PRL_1998,DellAnno_EPJS_2008,DellAnno_PRA_2007}.\\
Our proposed scheme would work best when the sender has some prior knowledge about the energy of the ensemble from which the input state is derived. This does not mean, however, that the sender has concrete information about the particular state to be teleported. If the sender has absolutely no information about the energy constraints on the input ensemble, the protocol would still function, but would yield suboptimal fidelities. In that case, the best strategy for the sender is to use a highly squeezed resource state so that the input states with large energies can be teleported with reasonably good fidelity. Before presenting all results, both in the noiseless and noisy scenarios, let us briefly discuss the classical limit for CV quantum teleportation which is essential to estimate any quantum advantage.

 %Note that these two distributions are qualitatively different. While the Gaussian distribution of inputs are product (independent) for the coherent amplitude and the squeezing strengths, the one with finite energy cutoffs ensures regularization in a correlated fashion. We would investigate their effects on the figures of merit in subsequent sections. But before that, let us briefly discuss the classical limit for CV quantum teleportation.
% {\color{red}{We will discuss later about these two types of input distributions elaborately.}}

\subsection{The classical limit}
\label{subsec:2a}

 In any quantum information protocol, it is necessary to define a classical threshold which quantifies the performance of the optimal classical routine for the task. If the figure of merit for the quantum protocol exceeds the classical limit, we can claim with certainty that quantum benefit is obtained. In quantum teleportation with
 %too, there exist classical limits which the fidelity must overcome to ensure that there is quantum advantage, for instance, in 
 the discrete qubit formalism, an average fidelity beyond $\mathcal{F} = 2/3$ indicates the presence of entanglement, thereby obtaining quantum advantage \cite{Horodecki_PRA_1999,Horodecki_PRL_1999,Linden_PRA_1999}.
 
 In CV teleportation, when the input states, say coherent states $\ket{\beta}$, are sampled from a Gaussian distribution $p(\ket{\beta}) = \frac{\lambda}{\pi}\exp{(-\lambda |\beta|^2)}$
 %are provided from a distribution which may be known to the two parties involved in the process 
 \cite{VanLoock_thesis_2002},   the optimal fidelity achievable through classical measure-prepare strategy is known to be $\mathcal{F}_{\mathrm{class}}^{\mathrm{coh}} \leq (1+\lambda)/(2+\lambda)$ \cite{Braunstein_JModOpt_2000}. If the distribution becomes completely flat, i.e. $\lambda = 0$, it reduces to $\mathcal{F}_{\mathrm{class}}^{\mathrm{coh}} \leq 0.5$. Therefore,  for states sampled from an infinitely flat distribution of energies, any fidelity above $0.5$ guarantees quantum advantage. However, if the standard deviation of a Gaussian distribution is finite ($\lambda > 0)$, or there exists a uniform distribution which contains states up to a particular energy only (as in Eq. \eqref{eq:p_uni}), the classical threshold increases beyond the aforementioned value. It can be intuitively understood since  it is easier for the concerned parties to replicate the input state through a measure-prepare strategy when  the states are drawn form a limited energy distribution \cite{Braunstein_JModOpt_2000}. Hence, the classical bound on the average fidelity depends on average input energy and decreases with the decrease in the spread of input energy.  Specifically, for a given distribution with a finite energy, we need to determine the corresponding fidelity which is achievable in  absence of entanglement. 
 
 Similarly, the optimal classical bound on the teleportation of squeezed states is not uniquely determined and
 %. It is now evident that the classical fidelity 
 is no longer bounded by $0.5$ \cite{Xiang_CPL_2005}. In Ref. \cite{Serafini_PRL_2007}, it was demonstrated that the classical protocol for sending squeezed states with a flatly distributed energy up to a maximum value can go higher than $0.9$. Similarly, for pure input squeezed states, a fidelity higher than $81.5\%$ is necessary to obtain quantum advantage, when the states belong to an infinite ensemble of uniformly distributed energy \cite{Adesso_PRL_2008}.

%\section{Motivation}
\label{sec:moti}
\begin{figure}
    \centering
    \includegraphics[width = \linewidth]{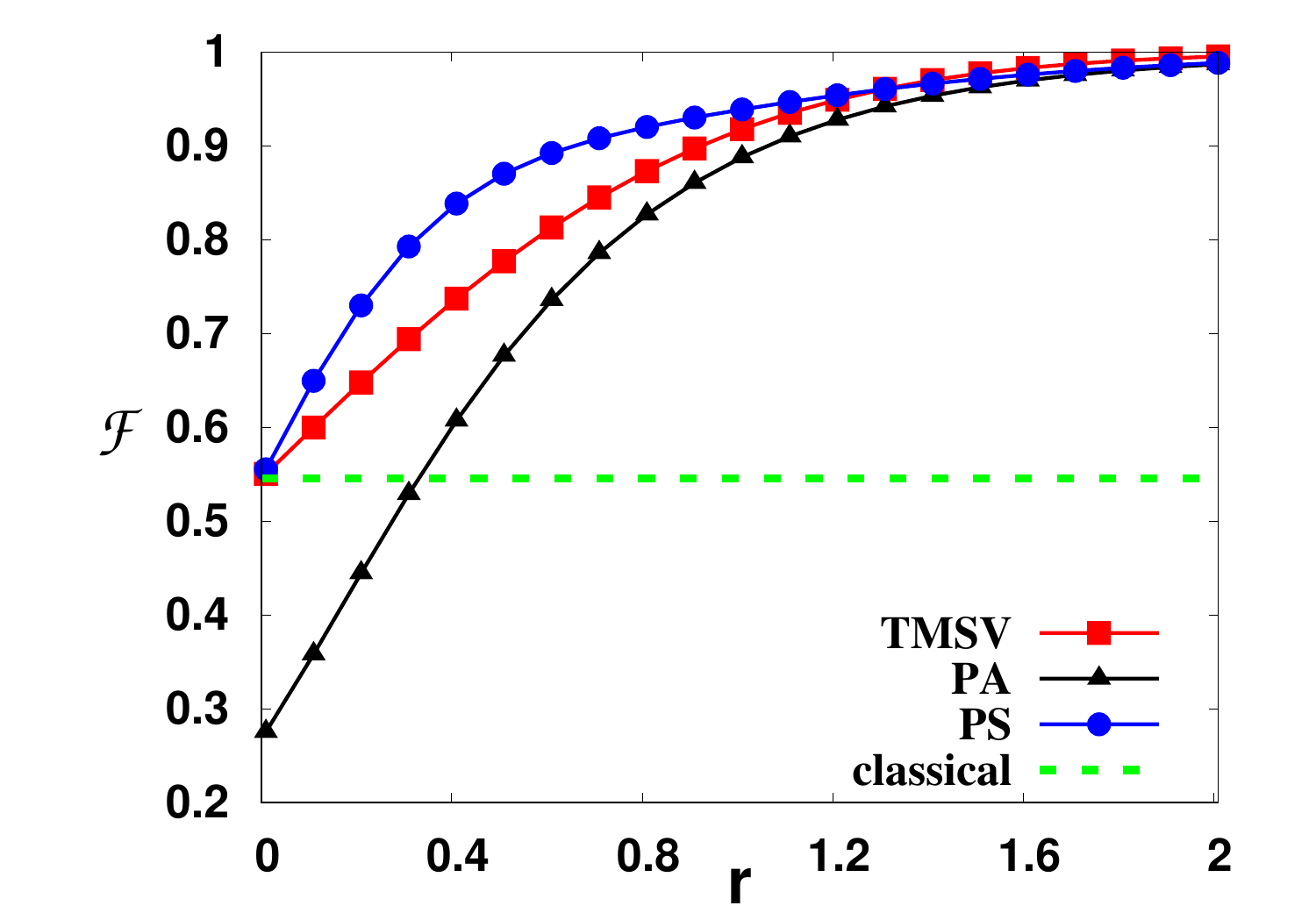}
    \caption{The average teleportation fidelity, \(\mathcal{F}\) (ordinate) of coherent states drawn from a Gaussian ensemble of variance, $\sigma = 5.0$ against the squeezing of the entangled resource states, $r$ (abscissa). Squares, circles,  and triangles represent the TMSV, photon added (PA) and  photon subtracted (PS) states as channels. The classical bound on the average fidelity is shown in dashed lines. It is interesting to observe whether such hierarchies among CV states change with different input distributions like uniform distribution with energy threshold.  Both the axes are dimensionless. }
    \label{fig:noiseless_avfid_r}
\end{figure}

\section{Characterizing  Noiseless CV teleportation via average Fidelity and fidelity deviation}
\label{sec:noiseless}

Before presenting the results in the absence of any kind of noise, let us specify the resource and input states considered here.

\textbf{Resources.} In our analysis, the shared resource state used  are squeezed Bell-like states which read as
\begin{eqnarray}
|\Phi\rangle = \hat{S}_{12}(\zeta) (\cos \delta|00\rangle + e^{i\eta} \sin \delta |11\rangle),
\label{eq:bell-state}
\end{eqnarray}
where $\hat{S}_{12}(\zeta) = e^{-\zeta a_1^\dagger a_2^\dagger + \zeta^\ast a_1 a_2}$ is the two-mode squeezing unitary operator with $\zeta=re^{i\gamma}$. It can be reduced to different well-known Gaussian and non-Gaussian states -- for $\delta=0$, it represents the two-mode squeezed vacuum state; choosing  $\delta=\arccos[(\cosh 2r)^{-1/2} \sinh r]$ and $\eta=\gamma-\pi$ gives the two mode photon added (PA) squeezed vacuum state; and  by choosing $\delta=\arccos[(\cosh 2r)^{-1/2} \cosh r]$ and $\eta=\gamma-\pi$, we obtain  the two mode photon subtracted (PS) state, where the last two are the non-Gaussian states. Note that a single photon is added (subtracted) in both the modes to create photon-added (photon-subtracted) states. In this work, comparative analysis of utility in using all three quantum resource states between the sender and the receiver is performed.

 \textbf{Inputs.} Three paradigmatic input states, namely the coherent state having displacement parameter $\beta = b e^{i \phi}$ given by $|\psi\rangle_{c} = \hat{D}(\beta) |0\rangle$, the squeezed state with squeezing parameter $\xi$ i.e. $|\psi\rangle_{s} = \hat{S}(\xi) |0\rangle$ and the squeezed-coherent state, $|\psi\rangle_{sc} = \hat{S}(\xi) \hat{D}(\beta) |0\rangle$ are considered for investigation. Here, $\hat{D}(\beta) = \exp(\beta \hat{a}^\dagger - \beta^* \hat{a})$ is the displacement operator and $\xi = \epsilon e^{i \theta}$. Notice that by examining the behavior of squeezed coherent states as inputs in QT,  the role of other input states on QT can be derived. 
 The analytical expression of the fidelity $f$ for teleporting a squeezed-coherent state using squeezed Bell state as a resource is given by \cite{DellAnno_PRA_2010_2}\\
\begin{multline}
     f_{|\psi\rangle_{sc}} = \frac{4}{\sqrt{\Lambda_1 \Lambda_2}}e^{\frac{\omega_1^2}{\Lambda_1} - \frac{\omega_2^2}{\Lambda_2}}\Big[ 1 + e^{-2r} \sin \delta (\Delta_2 \cos \delta - \Delta_1 \sin \delta) \\
     \left\{\frac{1}{\Lambda_1} \Big(1 + \frac{2 \omega_1^2}{\Lambda_1}\Big) + \frac{1}{\Lambda_2} \Big(1 - \frac{2 \omega_2^2}{\Lambda_2}\Big) \right\} \\ 
     + \frac{1}{4}e^{-4r} \Delta_2^2 \sin^2 \delta \Big( \frac{1}{\Lambda_1^2} \left\{ 3 + \frac{12 \omega_1^2}{\Lambda_1} + \frac{4 \omega_1^4}{\Lambda_1^2}\right\} \\
     + \frac{1}{\Lambda_2^2} \left\{3 - \frac{12 \omega_2^2}{\Lambda_2} + \frac{4 \omega_2^4}{\Lambda_2^2} \right\} \\
     + \frac{2}{\Lambda_1 \Lambda_2} \left\{ 1 + \frac{2 \omega_1^2}{\Lambda_1} - \frac{2 \omega_2^2}{\Lambda_2} - \frac{4 \omega_1^2 \omega_2^2}{\Lambda_1 \Lambda_2}  \right\} \Big) \Big],
     \label{eq:fid_squeezed_coh}
\end{multline}
   where the parameters $\Delta_1, \Delta_2, \Lambda_1, \Lambda_2, \omega_1^2$ and $\omega_2^2$ take  the form as
   \begin{eqnarray}
   && \Delta_1 = (1 + e^{4r}) + 2(1 - e^{4r})g + (1 + e^{4r})g^2, \nonumber\\
   && \Delta_2 = (1 - e^{4r}) + 2(1 + e^{4r})g + (1 - e^{4r})g^2, \nonumber\\
   && \Lambda_1 = e^{-2r} \Delta_1 + 2 e^{2\epsilon} (1 + g^2), \nonumber\\
   && \Lambda_2 = e^{-2r} \Delta_1 + 2 e^{-2\epsilon} (1 + g^2), \nonumber\\
   && \omega_1^2 = (1 - g)^2 (\beta - \beta^*)^2, \nonumber\\
   && \omega_2^2 = (1 - g)^2 (\beta + \beta^*)^2.
   \end{eqnarray}
   Here $g \in (0,1)$ is the gain factor involved in the measurement performed by the receiver \cite{VanLoock_thesis_2002}.
Equipped with this fidelity expression, we compute the maximal average fidelity ($\mathcal F$) by optimizing over $g$ and its corresponding fidelity deviation ($\Delta \mathcal{F}$)  both for the constrained uniform and Gaussian distributions of input states  using Eqs. \eqref{eq:av_fid} - \eqref{eq:p_gau}.
%\begin{equation}
%    \mathcal{F} = \frac{1}{\mathcal{N}} \int p(\beta,\xi) f_{|\psi_{\mathrm{in}}\rangle} d^2\xi d^2 \beta
%    \label{eq:av_fid_2}
%\end{equation}}
%\textcolor{blue}{1. alada kore ``numerically" lekhar dorkar nei,  
%~2. gaussian r g ta capital hbe throughout -- Gaussian, non-Gaussian  -- Sap}
 \begin{figure*}
    \centering
    \includegraphics[width = \linewidth]{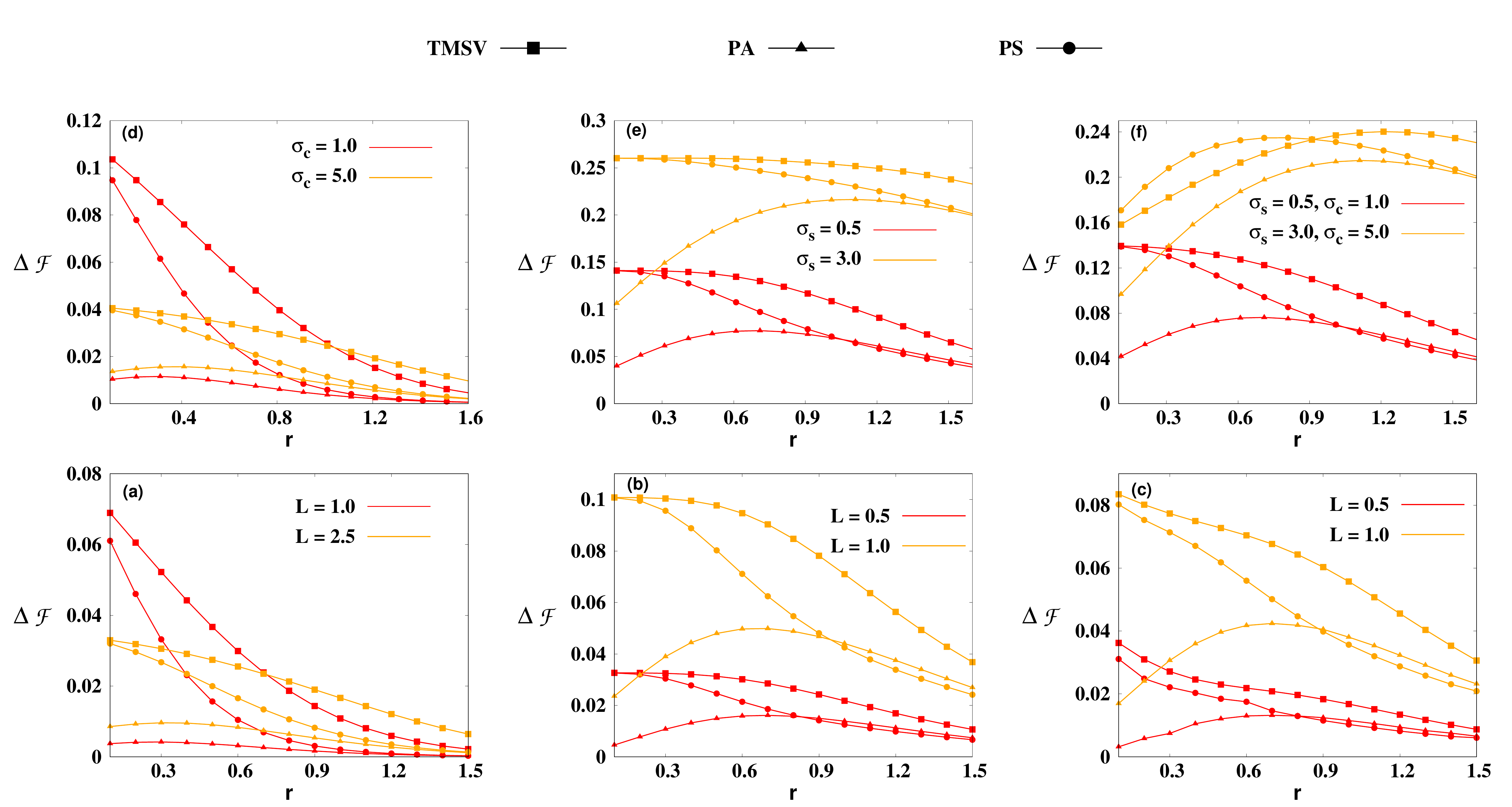}
    \caption{The variation of fidelity deviation, \(\Delta \mathcal{F}\) (vertical axis) vs. squeezing of the shared channels, \(r\) (horizontal axis) for uniform distribution (Bottom) and Gaussian distribution (Top) of different input states in the case of both Gaussian and non-Gaussian resource states. Symbols used for shared channels are same as in Fig. \ref{fig:noiseless_avfid_r}. Bottom: Plot of $\Delta \mathcal{F}$ for (a) coherent input states with energy cut-off $L = 1.0$ (dark (red)) and $L = 2.5$ (gray (yellow)), (b) squeezed input states with $L = 0.5$ (dark (red)) and $L = 1.0$ (gray (yellow))  and (c) squeezed-coherent input states with the same energy threshold specifications as in (b). 
    Top:  \(\Delta \mathcal{F}\) for different inputs, (d) coherent states  with $\sigma_c = 1.0$ (dark (red)) and $\sigma_c = 5.0$ (gray (yellow)), (e)  squeezed states with $\sigma_s = 0.5$ (dark (red)) and $\sigma_s = 3.0$ (gray (yellow)) and (f)  squeezed-coherent states having $\sigma_s = 0.5, \sigma_c = 1.0$ (dark (red)) and $\sigma_s = 3.0, \sigma_c = 5.0$ (gray (yellow)). All the axes are dimensionless.}
    \label{fig:noiseless_fidev_r}
\end{figure*}

\subsection{Trends of average fidelity and fidelity deviation with resource squeezing}

Let us  first investigate the response of the quality factors for teleportation with respect to the squeezing parameter, $r$ of the shared resource state. For all the three shared states considered here, namely the TMSV, PA and the PS states, the average fidelity increases monotonically with $r$ for both constrained uniform and Gaussian distribution of input states (see Fig. \ref{fig:noiseless_avfid_r}). This is intuitively satisfactory since the Einstein-Podolsky-Rosen (EPR) correlation increases with an increase of $r$ and the VBK protocol of teleportation uses EPR correlations as resource.  We will repeatedly return to these enhancement of features on increasing $r$ in situations where the average fidelity  fails to beat the classical limit.  
%Since it is well known that the average fidelity increases with the resource squeezing irrespective of any types of resource, we will not be discussing further about this behavior, and 
Instead of discussing the behavior of average fidelity which is  studied and known with \(r\), 
let us concentrate on the fidelity deviation with respect to   $r$  for different types of input states as well as resources and for a fixed average energy of the input distribution (see Fig. \ref{fig:noiseless_fidev_r}).  We categorize the trends according to the input states in the following manner. 
%Our main results are illustrated in for both the constrained uniform  as well as the Gaussian distributions.
%\subsubsection{The behaviour of fidelity deviation with resource squeezing}
% \label{subsubsec:gau_per_res_squeezing}

 \begin{description}
     \item[Squeezed states]  
     Unlike the average fidelity, a low value of fidelity deviation ensures  good performance of the resource states.  
     For input squeezed states, we observe that the photon added states provide the least deviation from the average fidelity for small resource squeezing, while the PS state accomplishes the task with minimum $\Delta \mathcal{F}$ for higher values of $r$. This is true when states are sampled both from the uniform (Fig. \ref{fig:noiseless_fidev_r} (b)) as well as the Gaussian distribution (Fig. \ref{fig:noiseless_fidev_r} (e)). On the other hand,  $\Delta \mathcal{F}$ increases with the increase in input energy, i.e.,  with the increase of \(L\) and \(\sigma_s\).  
     
     \item[Coherent states] We observe the decreasing trends of $\Delta \mathcal{F}$ with the increase of \(r\) in the resource, irrespective of the resource state. Like in the previous case,  PA states still provide the least deviation compared to PS or TMSV states  although the PS states overtake it  at a very high squeezing. Moreover, we find that unlike the squeezed states, there seems to be a complex relation between the squeezing in resource and energy threshold in inputs. In particular, $\Delta \mathcal{F}$  is low for ensembles with high energy up to a moderate value of $r$ both for the shared TMSV and PS states although the magnitude of the  squeezing required is more for the TMSV states than the PS states. For example, $\Delta \mathcal{F}_{L_\beta = 2.5} < \Delta \mathcal{F}_{L_\beta = 1.0}$ upto \(r_{PS} \leq 0.4\) while the similar hierarchy exists for the shared TMSV with a  higher \(r\), i.e.,  $\Delta \mathcal{F}_{L_\beta = 2.5} < \Delta \mathcal{F}_{L_\beta = 1.0}$ when \(r_{TMSV} \leq 0.7\). Similar behavior is also observed for the Gaussian distribution (as shown in Figs. \ref{fig:noiseless_fidev_r} (a) and (d)). 
     
     % \textcolor{red}{\sout{with the help of TMSV or PS states as resource,  The magnitude of the aforementioned resource squeezing is more for the TMSV states than the PS states --- $r_{TMSV} = 0.7$, while $r_{PS} = 0.4$ up to which $\Delta \mathcal{F}_{L_\beta = 2.5} < \Delta \mathcal{F}_{L_\beta = 1.0}$. For larger squeezing of the resource, $\Delta \mathcal{F}$ is lower for smaller input energy cut-offs. The PA state still provides the least deviation with the PS state overtaking it and providing the least deviation at very high squeezing.}}
      
     \item[Squeezed-coherent states]  The behavior of fidelity deviation with variation in resource squeezing for input squeezed-coherent states is similar to the other two inputs. The only significant difference is  the disparity in  \(\Delta \mathcal{F}\) for uniform and Gaussian distribution at higher energies. For states chosen from a Gaussian assemblage, the PS state constitutes the protocol with the highest value of $\Delta \mathcal{F}$ for low squeezing strengths at high input energies. As $r$ increases, its deviation falls below that of the Gaussian TMSV state (for $r \gtrsim 1.0 $) but still cannot overcome the one that is furnished by the PA states as resource. However, for constrained uniform distribution, the PS states teleport with minimum $\Delta \mathcal{F}$ at moderate to high \(r\) in the resource.
     \end{description}
 %\end{itemize*}

\subsection{Role of input energies in teleportation }
\begin{figure}
    \centering
    \includegraphics[width = \linewidth]{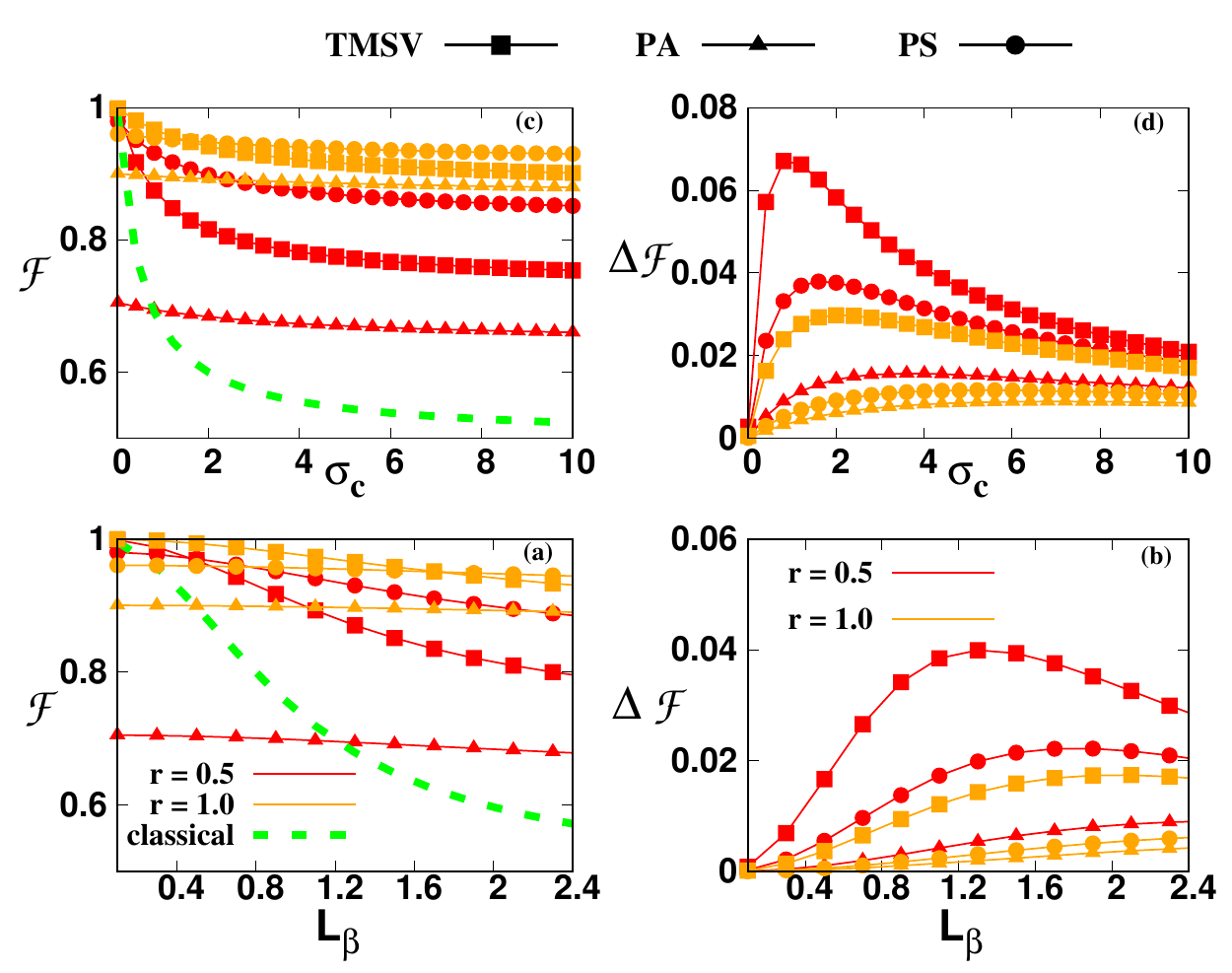}
    \caption{Average fidelity ((a) and (c)) and fidelity deviation ((b) and (d)) (ordinate) with respect to energy threshold, \(L_{\beta}\) in uniform distribution (Bottom) and \(\sigma_c\) for Gaussian distribution (Top) (abscissa). Symbols for channels are same as in Fig. \ref{fig:noiseless_avfid_r}. In (a)-(d),  inputs are taken to be coherent states having \(r=0.5\) (dark (red)) and \(1.0\) (gray (yellow)). 
    %Figure shows the variation of average fidelity  and the fidelity deviation against the energy cut-off for uniform distribution of input states in the case of both Gaussian and non-Gaussian resource states. (Bottom) $\mathcal{F}$ is plotted along the ordinate and the cut-off in energy is plotted along the abscissa for (a) coherent input states, (b) squeezed input states  and (c) squeezed-coherent input states. The TMSV state has been shown as squares, the PA state as triangles and the PS state as circles for resource squeezing $r = 0.5$ (in red) and $r = 1.0$ (in blue). The classical threshold in each case has been depicted as green dashed lines. (Top) Variation of $\Delta \mathcal{F}$ (ordinate) versus the energy cut-off (abscissa) has been shown for the exact same specifications as the Bottom.
    All the axes are dimensionless. }
    \label{fig:noiseless_avfidev_uni_energy}
\end{figure}

As mentioned before, one of the main focus of this work is to find the effects of the energy threshold in the input ensemble on the average fidelity and its deviation. Specifically, we examine   \(\mathcal{F}\) and \(\Delta \mathcal{F}\) with the variation of \(L\) in the constrained uniform distribution and \(\sigma_s\) as well as \(\sigma_c\) of the Gaussian distribution. 

%We now shift our attention to the case where we track the figures of merit of teleportation by varying the average energy of the input ensemble. Remember that the average input energy depends only on the standard deviation(s) of the gaussian distribution and the squeezing cut-off $L$ for the constrained uniform distribution. Also, note that the Gaussian distribution of the squeezed-coherent states has two independent standard deviations, $\sigma_s$ for squeezing parameter and $\sigma_c$ for displacement parameter respectively. \\

\begin{figure}
   \centering
   \includegraphics[width = \linewidth]{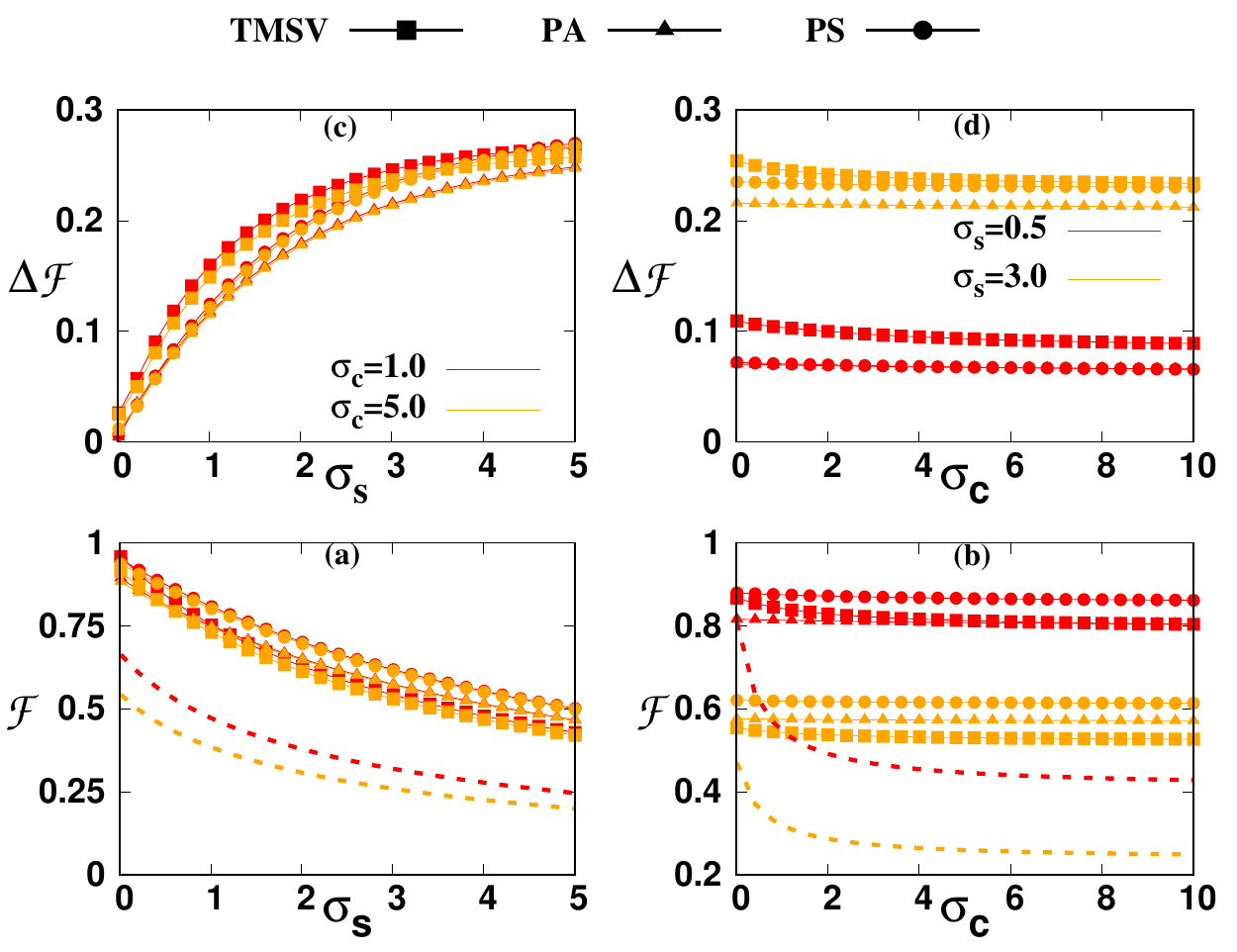}
   \caption{Average fidelity (Bottom) and fidelity deviation (Top) (ordinate) against the variance of squeezed-coherent input states, \(\sigma_s\) ((a) and (c)) and \(\sigma_c\) ((b) and (d)). 
   Symbols are same as in Fig. \ref{fig:noiseless_avfid_r} with resource squeezing $r=1.0$. %$\mathcal{F}$ (ordinate) is plotted against the standard deviation (abscissa) for (a) input squeezing parameter ($\sigma_s$) with 
   In (a) and (c), $\sigma_c=1.0$ (dark (red)) and $\sigma_c=5.0$ (gray (yellow)), while in (b) and (d),   $\sigma_s=0.5$ (in dark (red)) and $\sigma_s=3.0$ (in gray (yellow)).  The classical bounds on the average fidelity are shown in dashed line with respective colours. 
   %In Top, $\Delta\mathcal{F}$ (ordinate) is plotted against the standard deviations (abscissa) for the exact same specification as the Bottom.
   All the axes are dimensionless. } 
   \label{fig:noiseless_sqcoh_gau_energy}
\end{figure}

\begin{description}
   \item[Average fidelity] Let us illustrate the dependence of \(L\), \(\sigma_s\) and \(\sigma_c\) on \(\mathcal{F}\) for a fixed resource squeezing \(r\) which is chosen to be moderate (for demonstration, we choose e.g. \(r = 0.5\), and  \(1.0\)).  We observe that the average fidelity decreases monotonically with an increase in the input cut-off $L$ and with an increase in the variance $\sigma$ for a fixed value of  $r$ in the channel, irrespective of shared states and inputs as depicted   in Figs. \ref{fig:noiseless_avfidev_uni_energy} and \ref{fig:noiseless_sqcoh_gau_energy}. It is possibly due to the fact that the performance of QT decreases with the increase of energy to be teleported, indicated by the greater value of $L (\sigma)$.  Note, however, that a more involved picture emerges when inputs are drawn from the Gaussian distribution --   the rate of decrements in \(\mathcal{F}\) with respect to $\sigma_s$ is faster than that with $\sigma_c$ (see Fig. \ref{fig:noiseless_sqcoh_gau_energy}). We observe that to  transfer states with a high degree of squeezing or displacement, we require a highly squeezed resource state (containing high entanglement) to ensure that  the protocol is successful. We also find that the TMSV states can furnish a higher value of $\mathcal{F}$  for low energy Gaussian ensembles, with $\sigma \sim 0.1$ which depends also on the squeezing strength of the channel although PS states outperform over TMSV states in other ranges of input energies. 

  \item[Fidelity deviation] As seen in case of the average  fidelity, the increase of energy threshold in terms of increasing  \(L\) (\(\sigma\)) creates an obstacle in the success of the QT process, du to increase of the fidelity deviation with  energy, irrespective of the resource states and inputs, except for the  coherent states. In case of coherent states, \(\Delta \mathcal{F}\) exhibits a {\it nonmonotonic} behavior with input-energy, i.e.,  there is a threshold value of \(L\) and \(\sigma_c\) up to which it increases and subsequently decreases after the criticality. Such nonmonotonicity can be eliminated by increasing \(r\) of the channel (see Figs. \ref{fig:noiseless_avfidev_uni_energy} (b) and (d)). E.g. considering the TMSV state as resource, the criticality shifts from $\sigma_c \sim 0.8$ ($L \sim 1.2$) to $\sigma_c\sim 2.0$ ($L\sim2.0$) when the resource squeezing is increased from $0.5$ to $1.0$.  

\end{description}

From the patterns of \(\mathcal{F}\) and \(\Delta \mathcal{F}\), we can safely conclude that   for a teleportation protocol to succeed with a high average fidelity such that states are transferred with small variance in the desired fidelity, resource states with a moderate to high degree of squeezing are preferred, thereby demonstrating inverse proportionality between \(\mathcal{F}\) and \(\Delta \mathcal{F}\). 
In particular, by considering input squeezed states from Gaussian distribution,  the resource squeezing required to teleport states increases with the corresponding variance when the average fidelity is our major concern.
On the other hand,  in case of high average input energy, we need to make a compromise between the demand of high average fidelity and the low fidelity deviation in order to justify the quality of a resource state.

    \textbf{Quantum vs. Entanglement-free protocol. } Let us make a comparison between quantum protocols, which uses entangled channels, and entanglement-free (setting $r=0.0$) ones in terms of the  average fidelity. In this study, the squeezed or coherent states as inputs behave similarly compared to the squeezed coherent states.  For very low values of the variance, e.g. $\sigma_s\sim 0.2\text{ or }\sigma_c \sim 0.1$, or low input energy upper bound, $L \leq 0.1$,  with squeezed  or coherent states as inputs, the entanglement-free protocol performs equally well as the entangled one.   This may be due to the fact that for such low input energies, the unentangled protocol itself can furnish a very high average fidelity. As the energy of the input ensemble increases, the entanglement-based protocols win even with low values of the resource squeezing.
    However, such energy thresholds are not present in case of squeezed-coherent states as input, i.e.,   quantum routine outperforms the classical one in the entire range of both variances as shown in Fig. \ref{fig:noiseless_sqcoh_gau_energy}.
    Comparing resource states, we notice that for a uniform distribution in inputs, the TMSV and PS states always manage to beat the measure-prepare strategy while PA state can furnish quantum advantage only when the input energy is very high and the resource squeezing is substantial, say, $r \geq 1.0$.

\section{Resources hierarchies via fidelity deviation}
\label{sec:deviation}

In this section, we highlight situations where the average fidelity alone cannot completely characterize the performance in teleportation  by various resources. Specifically, we point out instances where 
resource states can be classified from 
%fidelity deviation can provide 
the non-trivial variations obtained in fidelity deviation. Moreover, our analysis reveals that several parameters like the squeezing of the resource, distribution of input states, and energy content play an important role in the performance of QT.  

%rank ordering better resource states for teleportation.
%To systematize our observations, before presenting our results with the most general input of squeezed coherent states, we focus on the special cases of coherent states and squeezed states as inputs.  

\subsubsection{Varying resource squeezing: Advantages of non-Gaussianity}

With respect to average fidelity alone, there is clear hierarchy of resource states with the PS being the best, closely followed by TMSV, while the photon-added states turn out to be the worst, failing to beat the classical limit in some cases. Let us now show that the ranking gets more involved if we take into account both the moments of the fidelity statistics. 

%However, when we factor into the information of fidelity deviation, rank ordering develops between resource states with respect to their performance in teleportation. 

When the input states are chosen to be squeezed or coherent states, for both constrained uniform and Gaussian distributions, we get qualitatively similar behavior of fidelity deviation.  The TMSV state shows the largest deviation among the three shared states. Therefore, for low $r$, PS is the best resource for quantum teleportation, since it not only possesses the highest average fidelity but also very low deviation, see Fig. \ref{fig:noiseless_avfidev_uni_energy}. For high values of $r$, the average fidelity for all the resource states grows, and become almost identical and therefore, the classification of resource states is entirely dictated by the fidelity deviation.  In this high $r$ limit, the deviation for PA and PS also become nearly equal while TMSV possesses a visibly larger deviation compared to these two. Therefore, here PA and PS become the better resource for quantum teleportation while TMSV turn out to be the worst. This feature also points out the role of non-Gaussianity in QT over Gaussian resources, especially for large squeezing.

For squeezed coherent inputs, things become more involved and we sometime get different responses for constrained uniform and Gaussian distributions. However, note that for low average energies of the input, it mimics a pattern similar to the previous cases. Things become interesting when relatively large values of  input  energies are considered. For example, for the Gaussian distribution,  the PS has a larger deviation compared to the TMSV state for a range of relatively low $r$ values.  This implies that for that range of $r$ values, we have to compare between two resources for which $\mathcal{F}_1> \mathcal{F}_2$ and also $\Delta\mathcal{F}_1 > \Delta \mathcal{F}_2$ are satisfied, see Fig. \ref{fig:noiseless_fidev_r}. Such a comparison of resources is not straight forward and  depends on the sensitivity requirements in deviation in a given context, see \cite{Saptarshi_PRA_2020}.

\subsubsection{Varying input energies}

%\begin{description}
\textbf{TMSV and PS as channels.} First of all,  TMSV  and  PS states can always beat the entanglement-free protocol provided that the squeezing is not too low and the input energy is moderately high. 
We observe that at a fixed squeezing strength of the resource states, the photon subtracted state accounts for a higher average fidelity than that of the TMSV state, when the input ensemble has a squeezing cut-off or variance over a certain value, viz. $L \gtrsim 0.8$ while the opposite hierarchy occurs in other situations.  For example, for  squeezed and coherent states as inputs belonging to a Gaussian ensemble up to a certain value of variance, e.g. $\sigma_s\leq 0.2$ and $\sigma_c\leq1.2$ (for $r=1.0$), the shared TMSV state between the sender and the receiver performs better than the others in terms of the average fidelity. Notice that such a ranking among states is not possible unless both fidelity and its deviation are taken in to account. 

%From Fig. \ref{fig:noiseless_avfidev_uni_energy} (a), (b) and (c), that the TMSV and the PS states have a point of crossover in the average fidelity graph for squeezed, coherent as well as squeezed-coherent states.For example, for input squeezed and coherent states belonging to a gaussian ensemble up to a certain value of standard deviation, e.g. $\sigma_s\leq0.2$ and $\sigma_c\leq1.2$ (for $r=1.0$), TMSV resource performs better than the others in terms of the average fidelity.  However, it should be noted that the critical value of the standard deviation of the input distribution up to which TMSV outperforms the others in terms of the average fidelity, increases with the resource squeezing $r$. In particular, this critical value of the standard deviation is quite sensitive to $r$ if we consider the coherent states as input. It is also interesting to note that in cases of squeezed states as input, TMSV beats PA as a resource in terms of average fidelity in the entire range of $\sigma_s$ for a lower value of resource squeezing $r$, e.g. $r=0.5$, while for higher values of $r$, e.g. $r=1.0$, situation becomes opposite. For input states with less energy, the gaussian TMSV state can furnish a better average fidelity than the non-Gaussian resources (see Bottom of Fig. \ref{fig:noiseless_avfidev_uni_energy}). With increasing resource squeezing the average fidelity becomes overwhelmingly better than the entanglement-free scheme even if the input squeezing cutoff is reasonably high. 

   \textbf{Photon added states.}  The fidelity deviation for the photon added state is very low, especially when we consider its variation with respect to $L$, and for a high value of $r$. The PA state, however, is not a suitable resource for QT, since it can only outperform the entanglement-free protocol once the squeezing is substantial. 
   %In fact, for input states of very low energy, the protocol with unentangled states can actually provide a better average fidelity than the photon added state upon optimization over the gain parameter. The only point of advantage for the PA state is that it allows teleportation of states with low fidelity deviation for both constrained uniform as well as gaussian distributions.

%\end{description}

The fidelity deviation helps removing the degeneracy among resource states in terms of being the optimal one in the teleportation protocol. We observe  that at high resource squeezing,
%all the resource states have almost the same average fidelity. Thus, 
according to $\mathcal{F}$, the non-Gaussian resources are always favorable over the TMSV one. 
%However, at the crossover point mentioned previously, it is not possible to select the optimal resource in this scenario. 
However, introducing  the fidelity deviation in picture, we find that only for high energy ensembles, the PS state offers the lower $\Delta \mathcal{F}$ along with high \(\mathcal{F}\),  thereby making it suitable for the QT purpose. 
%e the PS state for our teleportation purposes. 
Furthermore, at very low input energies, the average fidelity of the TMSV state is the highest among all states and the fidelity deviation, although higher than the non-Gaussian resources, is still very low ($O(10^{-2}))$, thereby making it  a reasonable resource as well. 
%the PS state offers the highest $\mathcal{F}$ and minimal $\Delta \mathcal{F}$. 
In the intermediate regimes, there is a competition between the high average fidelity offered by the PS state and low fidelity deviation  by the PA state although again the PS state is favorable due to high average fidelity leading to quantum advantage. The above discussion also manifests that although non-Gaussian resources can help to improve the teleportation protocol, the resource state must be chosen wisely, and also according to the input energy.

\section{Noisy CV teleportation}
\label{sec:noisy}

%\begin{figure*}
 %   \centering
  %  \includegraphics[width = \linewidth]{sq_noisy_fid.pdf}
   % \caption{Figure shows the average fidelity $\mathcal{F}$ (ordinate) against the noise parameters (abscissa) for TMSV (a) and (d), PA (b) and (e), PS (c) and (f) resource states with squeezed states as input. (Bottom) Average fidelity is plotted against the measurement noise $\mathcal{R}$ for $L = 1.0, r = 1.0$ (squares), $L = 1, r = 0.5$ (circles) and $L = 0.5, r = 1.0$ (triangles) at resource noise values $\tau = 0.0$ (in red) and $\tau = 0.3$ (green). We also depict the classical threshold corresponding to $L = 0.5$ (pink) and $L = 1.0$ (blue). (Top) Variation of average fidelity is shown against the resource noise $\tau$ for specific values of $\mathcal{R} = 0.0 ~ \text{(in red)} ~ \mathcal{R} = 0.2 ~ \text{(in green)}$ having the same other parameters as the Bottom.}
    %\label{fig:fig6}
%\end{figure*}

Upto now, the  investigations are carried out with the assumption that there is no noise in the preparation of resources or in the measurement process. Typically, imperfections are inevitable during the realization of these protocols
%the protocol cannot be realised ideally due to imperfections arising from  implementations 
in laboratories. In our analysis, we consider two main sources of noise - one occurring in the state itself, due to losses in the fiber through which the modes of the entangled resource are transmitted to the concerned parties, while  the other one arises  due to imperfect Bell measurements performed at the sender's node.

The noisy channel  quantified by $\tau = \gamma t$ is proportional to the fiber propagation length, where $\gamma$ is the mode damping rate \cite{DellAnno_PRA_2010,DellAnno_PRA_2010_2}, and the fiber loss factor is also associated with the interaction with a Gaussian bath of mean photon number $n_{th}$ which is taken to be zero in our work \cite{DellAnno_PRA_2010_2}. On the other hand, the imperfection in Bell measurement is considered by incorporating photon losses during the procedure which is modeled with the help of a beam splitter of transmittivity $\mathcal{T}$ and reflectivity $\mathcal{R}$. A non-zero value of $\mathcal{R}$ indicates finite losses in measurement  \cite{DellAnno_PRA_2010}. In the presence of the imperfections mentioned above, the expression of the one-shot fidelity for squeezed-coherent states can be written as
\begin{multline}
     f'_{|\psi\rangle_{sc}} = \frac{4}{\sqrt{\Lambda_1 \Lambda_2}}e^{\frac{\omega_1^2}{\Lambda_1} - \frac{\omega_2^2}{\Lambda_2}}\Big[ 1 + e^{-2r-\tau} \sin \delta \times ~~~~~~~~~~~~~~~~~~~~~~~~~ \\
     (\Delta_2 \cos \delta - \Delta_1 \sin \delta)\left\{\frac{1}{\Lambda_1} \Big(1 + \frac{2 \omega_1^2}{\Lambda_1}\Big) + \frac{1}{\Lambda_2} \Big(1 - \frac{2 \omega_2^2}{\Lambda_2}\Big) \right\} \\ 
     + \frac{1}{4}e^{-4r-2\tau} \Delta_2^2 \sin^2 \delta \Big( \frac{1}{\Lambda_1^2} \left\{ 3 + \frac{12 \omega_1^2}{\Lambda_1} + \frac{4 \omega_1^4}{\Lambda_1^2}\right\} \\
     + \frac{1}{\Lambda_2^2} \left\{3 - \frac{12 \omega_2^2}{\Lambda_2} + \frac{4 \omega_2^4}{\Lambda_2^2} \right\} \\
     + \frac{2}{\Lambda_1 \Lambda_2} \left\{ 1 + \frac{2 \omega_1^2}{\Lambda_1} - \frac{2 \omega_2^2}{\Lambda_2} - \frac{4 \omega_1^2 \omega_2^2}{\Lambda_1 \Lambda_2}  \right\} \Big) \Big],
     \label{eq:fid_squeezed_coh_noisy}
\end{multline}
	where
	%the parameters $\Delta_1, \Delta_2, \Lambda_1, \Lambda_2, \omega_1^2$ and $\omega_2^2$ having the following forms
   \begin{eqnarray}
   && \Delta_1 = (1 + e^{4r}) + 2e^{\tau/2}(1 - e^{4r})\tilde{g} + e^\tau(1 + e^{4r})\tilde{g}^2, ~~~ \nonumber \\
   && \Delta_2 = (1 - e^{4r}) + 2e^{\tau/2}(1 + e^{4r})\tilde{g}+ e^\tau(1 - e^{4r})\tilde{g}^2,~~~ \nonumber \\
   && \Lambda_1 = e^{-2r-\tau} \Delta_1 + 2 e^{2\epsilon} (1 + \tilde{g}^2) + 4 \Gamma, \nonumber \\
   && \Lambda_2 = e^{-2r-\tau} \Delta_1 + 2 e^{-2\epsilon} (1 + \tilde{g}^2) + 4\Gamma, \nonumber \\
   && \omega_1^2 = (1 - \tilde{g})^2 (\beta - \beta^*)^2, \nonumber \\
   && \omega_2^2 = (1 - \tilde{g})^2 (\beta + \beta^*)^2,
   \end{eqnarray}
with  $\tilde{g} = g\mathcal{T}$ and $\Gamma = \frac{1}{2}(1- e^{-\tau}) + g^2\mathcal{R}^2$ \cite{DellAnno_PRA_2010_2}. In presence of both the noises, the average fidelity and the fidelity deviation are calculated  after optimizing over $g$. %Again, we consider the TMSV state as the Gaussian resource and the photon added  and photon subtracted  states in the non-Gaussian sector. 
To study the effects of noise on QT, the moments of fidelity are studied with respect to a single noise parameter, while maintaining the other at a fixed value, for different regimes of resource squeezing and input energy. Moreover, to discuss the results systematically, our findings for the constrained uniform and Gaussian distribution of inputs are presented separately. \\
Notice that in discrete variable quantum teleportation, a possible noise model can involve contamination of the resource state with white noise \cite{Li_PRA_2021}. 
%In particular, they consider the qubit Werner state as the noisy shared state for teleportation. 
In the CV case, such admixing with white noise is unphysical since it corresponds to divergent energies. However, a CV version of the Werner state do exist as proposed in  Ref. \cite{Mista_PRA_2002} where a two mode squeezed vacuum state is admixed with a product of thermal states of identical temperatures (marginals of the initial TMSV). Although such states can be considered as resource, the computation of fidelity would be quite straightforward using linearity. Instead,
%motivated by the current literature of CV teleportation, we believe that the models of noise,
motivated by the noise models considered in Ref. \cite{DellAnno_PRA_2010_2}, which are argued to be close to  experiments, 
%.  Instead of modeling a noisy resource state through probabilistic admixture with white noise (or more physically the CV version of the Werner state),
we consider noise in various steps of the teleportation process and observe the effects on fidelity and its deviation. As mentioned before, it arises either from imperfect Bell measurements by the receiver (quantified by $\mathcal{R}$) or when the resource is distributed between the two parties through a fiber channel (quantified by $\tau$).  The noise models considered in our paper are rooted in the experimental implementation of CV teleportation and our analysis will shed light on how such noise can be tackled by adjusting the other parameters in the system.

 \subsection{Average fidelity  with constrained uniform input distribution: Gaussian resources are better}
 \label{subsec:av_fid_noisy}

 \begin{figure*}
    \centering
    \includegraphics[width = \linewidth]{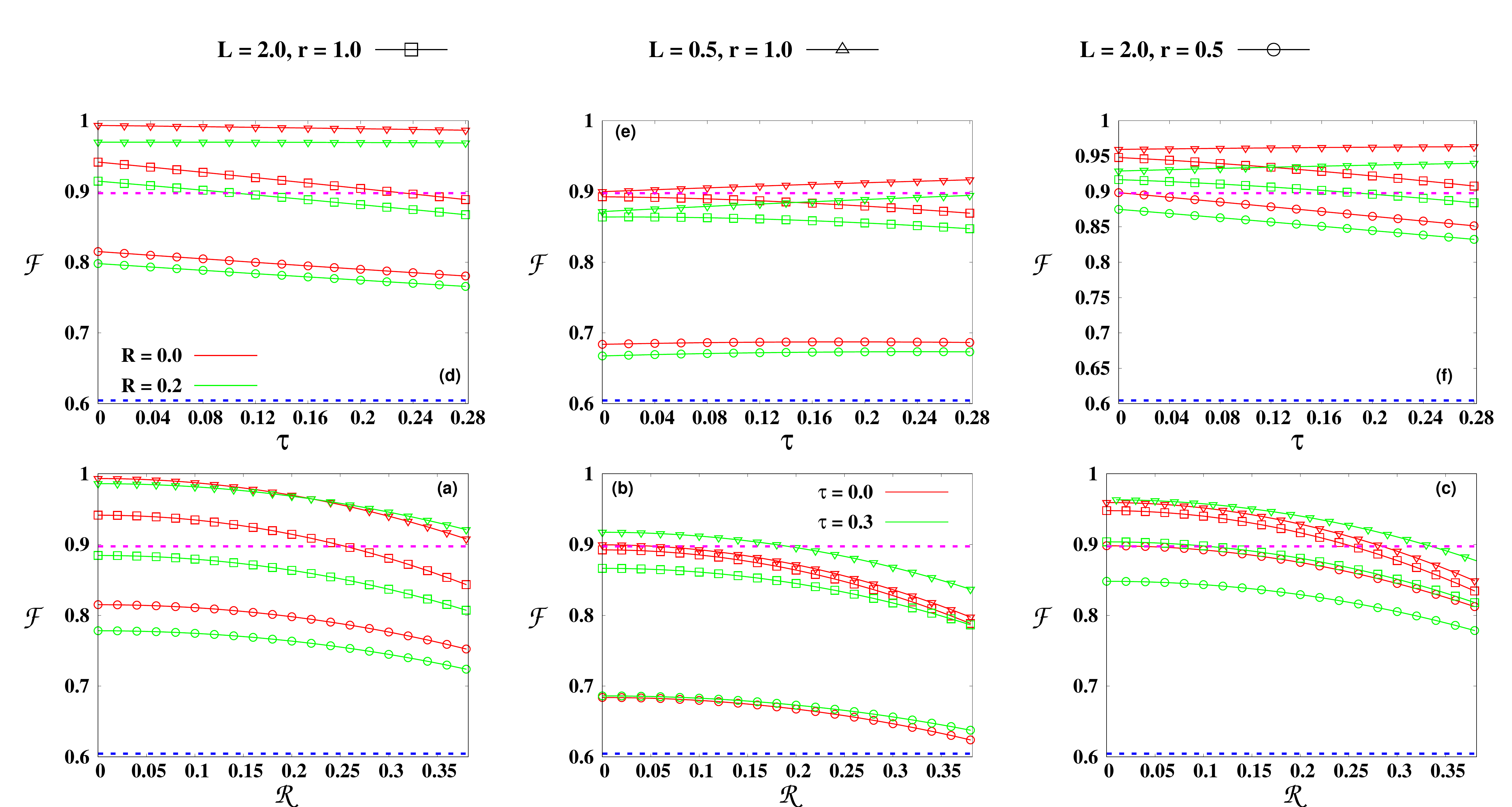}
    \caption{Average fidelity (along ordinate)  vs. noise parameters, \(\mathcal{R}\) (Bottom) and \(\tau\) (Top)  (along abscissa)  for TMSV (a) and (d), PA (b) and (e), PS (c) and (f) resource states with coherent states as input. (Bottom)  $\mathcal{R}$ is varied for fixed $L = 2.0, r = 1.0$ (open squares), $L = 0.5, r = 1.0$ (open triangles) and $L = 2.0, r = 0.5$ (open circles) at resource noise values $\tau = 0.0$ (dark (red)) and $\tau = 0.3$ (gray (green)). The classical bounds obtained by measure-prepare strategy are shown  corresponding to $L = 0.5$ (dashed gray (pink)) and $L = 2.0$ (dashed dark (blue)). (Top) When noise, $\tau$ in the channel varies,  the measurement noise are fixed to  $\mathcal{R} = 0.0$ (dark(red))~ $\mathcal{R} = 0.2$ ~ (gray (green)).  All other specifications are same as the Bottom. All the axes are dimensionless. }
    \label{fig:fig7}
\end{figure*}

 Let us first consider the variation of the average fidelity $\mathcal{F}$ with the measurement noise $\mathcal{R}$, for fixed values of the noise in channels, $\tau$. As expected, the average fidelity decreases with an increase in the magnitude of $\mathcal{R}$, which is illustrated in the Bottoms of Fig.  \ref{fig:fig7}.  However, it can be increased if the resource squeezing is high or the input energy is low. Contrary to the noiseless scenario, the PS state provides the highest $\mathcal{F}$ only for high energy input ensembles ($L_\xi = 1.0$) with low resource squeezing strengths  and low noise limits up to $\mathcal{R}, \tau = 0.1$. Otherwise, when the states to be teleported are of high energy and the resource squeezing is also sufficient, \(\mathcal{F}\) for the Gaussian TMSV state is slightly higher than that of the PS state, 
 %proves to be the optimum resource for noisy teleportation
 %This is mainly due to the fact that the TMSV state is 
 thereby indicating its robustness against  noise and also proving its appropriateness for noisy CV teleportation. 
 %provided 
 %the states to be teleported are of high energy and the resource squeezing is also sufficient. 
 %and can sustain a reasonable average fidelity over the entire domain of parameter values. 
 %Our results are an important benchmark in the noisy scenario, since it seems that Gaussian states are the good resource, provided that the states to be teleported are of high energy and the resource squeezing is also sufficient. This is distinctly different from the ideal \textcolor{blue}{noiseless} protocol, where non-Gaussian states always manage to outperform the TMSV state in regimes of high resource squeezing. 
 %Note, however,  the 
 %PS state provides average fidelity only slightly lower than its Gaussian counterpart, and for low energies, it again establishes its dominance in the teleportation protocol (see Fig. \ref{fig:fig6}).
 
 \subsubsection{Counteracting one noise with the other - a constructive effect} 
 
Let us report here an interesting feature  when different values of the resource squeezing are considered. By varying \(\mathcal{R}\), one would expect the average fidelity to be low for higher values of $\tau$, i.e., in presence of both the noises. This is indeed the case but not over the entire range of $\mathcal{R}$. We observe
%in Figs. \ref{fig:fig6}(a) and \ref{fig:fig7}(a), 
that there exists a region in $\mathcal{R}$ where  \(\mathcal{F}\)  is higher in presence of resource noise, say $\tau = 0.3$ than that of the scenario without noise in resources, i.e. with \(\tau =0\). 
%when $\mathcal{R} \in (0,0.4)$: what do you mean $\mathcal{R} > 0.4$?}, the average fidelity is more for a higher value of the resource noise ($\tau = 0.3$). This 
This can be interpreted as if the effect of one kind of noise is countered by the other one, thereby exhibiting a {\it constructive phenomenon} which is more pronounced in case of coherent inputs (see Fig. \ref{fig:fig7}). 
It may also indicate that when the resource is affected by ineffective propagation, the protocol may not be the optimal even with a properly tuned gain parameter $g$. The point of crossover depends on the resource squeezing as well as on the energy of the input state.  Comparing   Fig. \ref{fig:fig7} (a) with (b) and (c), we realize that the constructive effect is more visible for non-Gaussian states compared to the Gaussian ones. Notice that the advantage is counted only when \(\mathcal{F}\) obtained in a noisy scenario is higher than the entanglement-free protocol without noise which we will discuss later.

%For lower values of the input energy cut-off, $L = 0.5$, the average fidelity corresponding to higher resource noise  is better at a smaller value of $\mathcal{R}$ (e.g. $\mathcal{R} \sim 0.15 $), whereas, as $L$ increases to $1.0$ , the measurement noise at which this constructive effect is exhibited also increases to $\mathcal{R} \sim 0.25$. This is especially evident when the input state is drawn from a squeezed ensemble. For coherent and squeezed-coherent states, this feature is exhibited by the Gaussian resource state for low input energies. In case of the non-Gaussian PA and PS states, the constructive outcome is even more prominent with higher average fidelity always resulting for higher values of $\tau$ for sufficiently large squeezing strengths $r \geq 1.0$ (see Figs. \ref{fig:fig7} (b) and (c)). Thus for coherent and squeezed-coherent input states, the non-Gaussian resources pose an advantage over the Gaussian TMSV state in the noisy setting.\\

% \subsubsection{Effects of measurement noise on different resources}
The effects of noise on the average fidelity is also distinctive for different classes of input states. In particular, \(\mathcal{F}\) for the squeezed-coherent states ($\mathcal{F} \geq 0.4$) is much lower than that of the squeezed and coherent inputs ($\mathcal{F} \geq 0.7$), especially for high ensemble energies, which is not the case in the noiseless scenario. Moreover, the average fidelity decreases at a much faster rate for the PS and PA states, which indicates that the impact of noise is more on non Gaussian states in comparison with TMSV state having moderate squeezing. 
%One possible reason for obtaining the better fidelity with the TMSV state when the squeezing available is sufficient, thereby again proving its robustness.
Furthermore, the difference between $\mathcal{F}$ at higher and lower values of $\tau$ is least for the PA state, $\sim 0.005$, but significantly more for the TMSV and PS states, $\sim 0.01$. 
%This implies that photon addition makes the resource state much less susceptible to noise. Thus, in the noisy non-ideal measurement protocol, the TMSV state proves to be the most effective resource in terms of the average fidelity. 

 \subsubsection{Robustness against resource noise}

 Let us now fix a moderate amount of noise in measurement (e.g. we choose $\mathcal{R} = 0.0$, and  $0.2$) and study 
 the dynamics of average fidelity by varying  noise  in the shared channel. First of all, no constructive effects with \(\tau\) is seen by comparing \(\mathcal{R} =0\) and \(\mathcal{R} = 0.2\) (see Fig. \ref{fig:fig7}).  However,  
 the decrease in $\mathcal{F}$ with increase of $\tau$ is much slower than the one observed by varying $\mathcal{R}$ especially when  the squeezing strength in resource is high,  irrespective of  Gaussian or non Gaussian resource states and inputs (comparing upper and Bottom of Fig. \ref{fig:fig7}). It demonstrates the adverse effects of inefficient measurement on the protocol compared to noise in resource states. However, such detrimental impact can again be wiped out in presence of high squeezing in the shared channel. 
 
 %is not as rich as in the previous case. Our observations are illustrated in Tops of Fig. \ref{fig:fig6} and Fig. \ref{fig:fig7}.  The average fidelity for higher measurement noise values is less than for lower values and there are no crossovers indicating \textcolor{blue}{there are no} constructive effect. On the other hand, the decrease in $\mathcal{F}$ with increase in $\tau$ is much slower than with increase in $\mathcal{R}$. At higher resource squeezing the decline in average fidelity is at a very low rate. This implies that the teleportation protocol is affected much less by noises in the resource \textcolor{blue}{than in the} transfer process. Once more, the TMSV state emerges as the most valuable resource while the PA state, although having the least variation in $\mathcal{F}$, turns out to be worse than the Gaussian TMSV and the non-Gaussian PS  states. This fact has already been hinted at in the previous sections. The average fidelity for  coherent states is the highest among the input ensembles which shows that they are least affected by noises in the resource.

 \subsubsection{Comparison with the unentangled protocol}
 \label{subsubsec:noi_uni_classical}

The teleportation protocol with unentangled states (classical protocol) involves a measure-prepare routine, which evidently does not suffer from the noise models considered here. Therefore, it is justified to examine whether the noisy teleportation process can beat the noiseless classical one. 

All the different scenarios have so far been compared keeping in mind the quantum advantage, i.e., the shared TMSV states with coherent inputs to be teleported  exhibits maximum robustness against both the noise models considered here. Moreover, as the input energy increases, the TMSV state can retain quantum advantage in presence of large amount of noise. For example, for $L = 0.5$, the TMSV state with $r = 1.0$ can outperform the unentangled protocol up to $\mathcal{R} \sim 0.16$, while   the same resource can retain quantum advantage for $\mathcal{R} \leq 0.28$ with $L = 1.0$ in case of squeezed input states . The situation changes  in case of the squeezed-coherent ensembles when the TMSV state can outperform the classical scheme only for low input energies  and for higher values of $L$ only up to small  magnitudes of noise.

%Towards that end, we describe how different resource states compare in the teleportation protocol with respect to the unentangled threshold.

%We have already established that the Gaussian TMSV state is more robust than the PA and PS states and offers a higher average fidelity in the noisy scenario. As such, we observe that quantum advantage is provided by the TMSV state for high values of the resource squeezing and low noise parameters in case of squeezed input states with high energy cut-off. This is because, for large input energies, the classical protocol cannot do well in transferring the state and the TMSV state helps in overpowering the measure-prepare strategy. Moreover, as the input energy increases, the TMSV state can retain quantum advantage for a larger range of the noise. For example, for $L = 0.5$, the TMSV state with $r = 1.0$ can outperform the unentangled protocol up to $\mathcal{R} \sim 0.16$, while for $L = 1.0$, the same resource can retain quantum advantage for $\mathcal{R} \leq 0.28$ as shown in Fig. \ref{fig:fig6}(a). For coherent states, which are most robust against noise, the Gaussian state is beneficial for both low and high values of the input cut-off energy over the entire range of the noise parameters $\mathcal{R}$ and $\tau$ (see panels (a) and (d) of Figs. \ref{fig:fig6} and \ref{fig:fig7}). The situation changes dramatically in case of the squeezed-coherent input ensemble when the TMSV state can outperform the classical approach for low input energies and for higher values of $L$, it can sustain advantage only up to small noise magnitudes.

In case of  photon added state, the regimes of quantum advantage are very limited especially for squeezed input states and for low input energy.  Quantum advantage can only be found for low \(\mathcal{R}\) and \(\tau\). 
%For input squeezed states, the PA state can outperform the classical procedure only for high input energies ($L = 1.0$) with high resource squeezing ($r \geq 1.0$) when the noise parameters assume low values ($\mathcal{R}, \tau \leq 0.08$). It can however, always perform admirably in case of coherent states drawn form a high energy ensemble. For low $L_\beta$, gain against the classical scheme is available only for low $\mathcal{R}$ and $\tau$ as shown in Figs. \ref{fig:fig7} (b) and (e). The situation is similar for squeezed-coherent states in case the non-Gaussian PA state has sufficient squeezing $r \geq 1.0$.
The PS state performs better than the its photon added counterpart irrespective of inputs. Again, 
it performs best for coherent input states, always outperforming the classical measure-prepare routine for high input energy. For low $L_\beta$, it can furnish quantum advantage with sufficient squeezing ($r \geq 1$) unless $\mathcal{R}$ and $\tau$ are too high while both for squeezed  and squeezed-coherent input states, the entangled states win over the classical protocol with high  squeezing  and energy cut-offs when noise in the channel and measurements is low.  

%For squeezed input states, it can retain quantum advantage for larger regimes of noise when the input energy is high (e.g. $\mathcal{F}$ is higher than that of the classical protocol up to $\mathcal{R} \sim 0.12$ for $L_\xi = 1.0$), while no advantage is obtained for low squeezing cut-off (see Figs. \ref{fig:fig6} (c) and (f)). Again, 
%it performs best for coherent input states, always outperforming the classical measure-prepare routine for high input energy. For low $L_\beta$, it can furnish quantum advantage with sufficient squeezing ($r \geq 1$) unless $\mathcal{R}$ and $\tau$ are too high. When teleporting high energy squeezed-coherent states, it beats the unentangled protocol only when the squeezing is high and the noise is low. For lower energy input distributions, it can perform better when one noise is countered by the other, e.g. when $\tau = 0.3$ can cancel effects of the measurement noise $\mathcal{R}$.

 \subsection{Effects of  Gaussian input distribution on noisy teleportation}
  \label{subsec:av_fid_noisy_gau}
  
 A similar examination is carried out when the input states are sampled from a Gaussian distribution. Unlike the constrained uniform distribution, all input states, squeezed, coherent and squeezed-coherent  share more or less similar properties  of average fidelity with respect to both noise parameters. So, we mainly discuss the behavior of average fidelity for input squeezed states and explicitly mention the corresponding situations for other input states whenever we come across any  individual feature.
 
   As already emphasized, we will only present those  situations in which the performance of QT protocol is better than the prepare-measure strategy even in the presence of noise. 
   %Now we are going to examine whether the noisy quantum teleportation protocol can outperform the noiseless classical one. 
   Like in the uniform case,  the quantum process always outperforms the classical one in case of teleporting input coherent states irrespective of all resources and noise models that are considered here. However,  with different choices of average input energy and resource squeezing, the average fidelity may get affected differently. Nevertheless, as in the noiseless scenario, we can get a quantum advantage over the classical one when the input energy or resource squeezing is reasonably high. 
   %Even for squeezed and squeezed-coherent input states, the aforementioned condition is applicable to achieve quantum advantage.
 
%  

 %\textcolor{red}{not checked from here till next section by Aditi.}
 %In the present section, we will study the variation of average fidelity $\mathcal{F}$ with the noise parameters $\mathcal{R}$ and $\tau$ when the input states are sampled from a Gaussian distribution. Here we notice that squeezed, coherent or squeezed-coherent - all input states share more or less similar properties of variation of average fidelity with respect to both noise parameters. So, we mainly discuss the behaviour of average fidelity for input squeezed states and explicitly mention the corresponding situations for other input states whenever we come across any discrepancy or individual feature.
 
% \subsubsection{Effects of measurement noise over ideal teleportation}
 
     %In Bottom of Fig. \ref{fig:fig9}, we plot the behaviour of average fidelity with respect to measurement noise for coherent input states while considering a fixed resource noise. We study our observations for different types of resources- TMSV(left), PA(middle) and PS(right).
     
     \begin{figure}
    \centering
    \includegraphics[width = \linewidth]{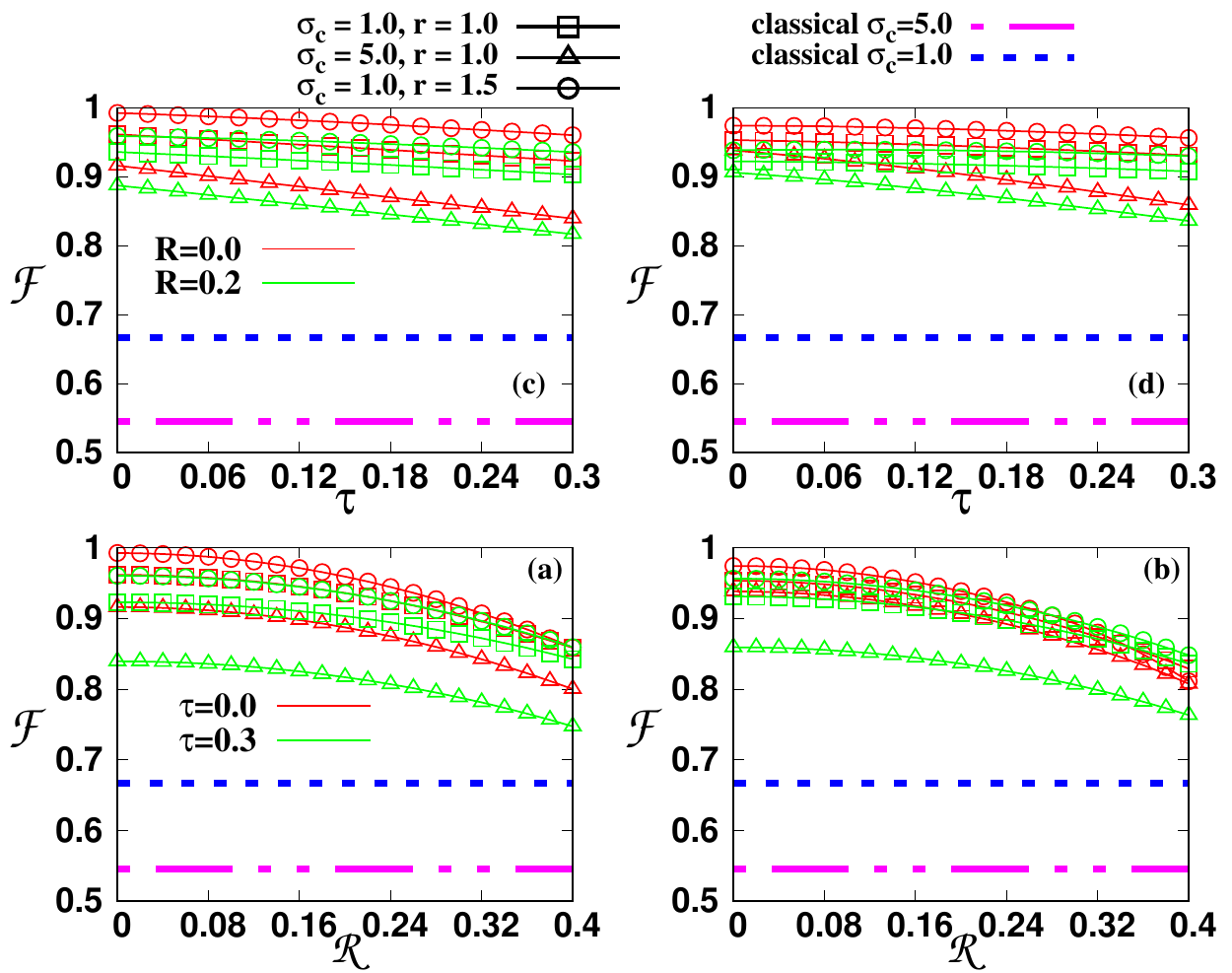}
    \caption{Average fidelity  (ordinate) with noise parameters, \(\mathcal{R}\) (Bottom)  and \(\tau\) (Top) (abscissa). Input states are chosen to be again coherent states drawn from a Gaussian distribution with TMSV (left panel)
     and PS (right panel) as resources. %(Bottom) $\mathcal{F}$ against the measurement noise $\mathcal{R}$ at resource noise values of $\tau = 0.0$ (in dark (red)) and $\tau = 0.3$ (in gray (green)) for TMSV and PS states in (a) and (b) respectively. (Top) Variation of $\mathcal{F}$ with resource noise $\tau$ is illustrated at measurement noise $\mathcal{R} = 0.0$ (in dark (red)) and $\mathcal{R} = 0.2$ (in gray (green)) for (c) TMSV and (d) PS states. 
    In all the panels, the variance and resource squeezing are depicted as $\sigma_c = 1.0, r = 1.0$ (open squares), $\sigma_c = 5.0, r = 1.0$ (open triangles) and $\sigma_c = 1.0, r = 1.5$ (open circles). The classical threshold corresponding to $\sigma_c = 5.0$ is shown with dot-dot-dashed gray (pink) lines while the dotted dark (blue) line represents $\sigma_c = 1.0$. All other specifications are same as in Fig. \ref{fig:fig7}. All  axes are dimensionless.}
    \label{fig:fig9}
\end{figure}
     
{\bf Impact of measurement noise on average fidelity.}     We observe that the average fidelity decreases monotonically with   $\mathcal{R}$, for all types of resources and as well as inputs. This is quite expected since in general noise causes some hindrance in any protocol. However, three interesting features emerge which are discussed as follows (see Fig. \ref{fig:fig9}). 
    % \begin{figure*}
%    \centering
%    \includegraphics[width = \linewidth]{avgfid_noisy_sqz_gau.pdf}
%    \caption{for squeezed input, TMSV(left),PA(middle),PS(right)}
%    \label{fig:fig10}
%\end{figure*}
% \begin{figure*}
 %   \centering
  %  \includegraphics[width = \linewidth]{avgfid_noisy_sqzcoh_gau.pdf}
   % \caption{for squeezed-coherent input, TMSV(left),PA(middle),PS(right)}
%    \label{fig:fig11}
%\end{figure*}
     \begin{itemize}
     
         \item 
         %Considering different resources we observe that for squeezed states or coherent states as input, 
        Considering only \(\mathcal{F}\), Gaussian shared channels are again more robust against measurement noise in absence of resource noise, i.e.,  \(\tau =0\) as compared to non-Gaussian ones. 
         %there is no noise in resources. 
         More precisely, for coherent states as inputs, the difference between \(\mathcal{F}\) at low and high values of measurement noise  is prominent for low input average energy and high resource squeezing while for squeezed-coherent state, this feature is noticeable at high average energy of the input. E.g. when $\sigma_c=1.0$ and $r=1.5$, we define $\delta\mathcal{F}=  \mathcal{F}^{\mathcal{R}=0.0} - \mathcal{F}^{\mathcal{R}=4.0}$ which for TMSV is  $\delta\mathcal{F}_{TMSV}=   0.133$, for PA is $\delta\mathcal{F}_{PA}=0.165$ and for PS,  $\delta\mathcal{F}_{PS}=0.163$. 
         %where $\delta\mathcal{F} = \mathcal{F}^{\mathcal{R}=0.0} - \mathcal{F}^{\mathcal{R}=4.0} $ values are calculated by taking the differences of average fidelity at $\mathcal{R}=0.0$ and $\mathcal{R}=0.4$. \\
         
         Typically, we expect that to obtain a better average fidelity with a fixed average energy, we require a resource with high squeezing. However,  we observe 
         %that beyond a certain value of $\mathcal{R}$ for a particular input distribution and resource type, the trend is opposite. This feature is observed for 
         an opposite behavior with coherent states as inputs and  PS states as quantum channel  in presence of a noise  only in measurements (i.e., taking $\tau=0.0$). For example, 
         %using PS state as a resource for input coherent states with %
         with $\sigma_c=1.0$,  beyond $\mathcal{R}\sim 0.35$, the low resource squeezing helps to manage better average fidelity than the states with high squeezing.  A similar trend is also observed in case of teleporting squeezed input states with TMSV and PS resource although the quantum advantage is unattainable there.
         
         \item Another point of interest is the `constructive effect' of noise that we have already noted in  case of uniform distribution of the input states. 
         %Here also we get the same behaviour, i.e., the value of $\mathcal{R}$ beyond which this constructive effect is observed depends on both the resource squeezing as well as the standard deviation of the distribution. We see that this constructive effect is more prominent for lower value of the input average energy (i.e., the standard deviation)  as well as resource squeezing $r$ irrespective of any resources as well as input states considered. E.g., considering coherent input states with PA resource state, we observe that the value of $\mathcal{R}$ for $r=1.0$ and $1.5$ is found to be $\mathcal{R}\sim0.18$ and $\mathcal{R}\sim0.26$ respectively when the average input energy is as low as $\sigma_c\sim1.0$. 
         Moreover, as in the constrained uniform distribution, it is noted that the constructive effect starts at relatively lower values of $\mathcal{R}$ for non-Gaussian resources compared to the Gaussian ones irrespective of squeezed or coherent states as inputs. E.g. considering $r=1.5$ and $\sigma_c=1.0$,  constructive effect emerges with the measurement noise values for different shared states as $\mathcal{R}_{TMSV}\sim0.40$, $\mathcal{R}_{PA}\sim0.26$, $\mathcal{R}_{PS}\sim0.28$.

         %
         %This behaviour is also true for input squeezed states, but the quantum advantage is beyond our reach as we mentioned earlier.
         
         \item Resources with lower squeezing strength corresponding to a fixed input distribution are less sensitive against measurement noise. This feature can easily be noted from  Fig. \ref{fig:fig9} when we compare $r=1.0$ and $1.5$ for a fixed variance. Moreover,  noisy channels  help to retain the robustness against the noise in measurement. 
     \end{itemize}

     %It is interesting to note that for higher values of measurement noise (considering $\tau=0.0$), TMSV outperforms PS in terms of average fidelity when squeezed states are considered as input, which is exactly opposite to the scenario that we saw in noiseless case. However, considering coherent states as input we observe that the different types of resources follow more or less similar properties as we saw in noiseless case.

     %\subsubsection{Behaviour against resource noise}
     {\bf Response against resource noise.} As seen in case of constrained uniform distribution, \(\mathcal{F}\) exhibits some distinct features  in this noise model which are either not observed or not pronounced in presence of noise in measurements.

    % In Top of Fig. \ref{fig:fig9}, we depict the behaviour of average fidelity with respect to resource noise at a fixed measurement noise for coherent states as input. We study our observations for different types of resources- TMSV(left), PA(middle) and PS(right). The behaviour of average fidelity with respect to varying resource noise is not that much interesting here. The average fidelity decreases with the increasing resource noise irrespective of all types of inputs as well as resources as it is expected. A a few points can be considered -
     \begin{itemize}
         \item Greater average energy of the input states makes the performance of the protocol less robust against resource noise. The measurement noise slightly improves the robustness of the performance against the resource noise.
         
         \item We can see that with small values of $\sigma_c$ ($\sim1.0)$ for coherent states as input, non-Gaussian resources are more robust than the Gaussian one against resource noise (see Fig. \ref{fig:fig9})
         %. On the other hand, 
         while Gaussian states are the best for input squeezed and squeezed-coherent states
         %, we notice that the Gaussian resources are more robust %
        in the high input energy regime.
        
        % \item The behavior of average fidelity with respect to different resources shows overall similar properties that we have already seen in noiseless scenario. Sometimes a higher value of resource noise helps to improve the qualitative status of TMSV state as a resource when comparing to PS state in the case of squeezed and coherent (for high resource squeezing in case of input coherent states) states as input. This may be due to the fact that TMSV state is more robust against resource noise comparing to others for that particular input and resource condition. However, for squeezed-coherent states as input, we observe that PS always manages to provide the better average fidelity in the teleportation protocol.
         
     \end{itemize}
     
     Summarizing, we find that in a noisy scenario,  both TMSV and PS states are good quantum channels for QT according to the average fidelity regardless of energy distribution of inputs and input states. It will now be interesting to enquire whether the patterns of fidelity deviation can help us to identify the suitable  resource  for QT.

\section{Role of Fidelity deviation for noisy teleportation}
\label{sec:fid_dev_noisy}

 \begin{figure}
    \centering
    \includegraphics[width = \linewidth]{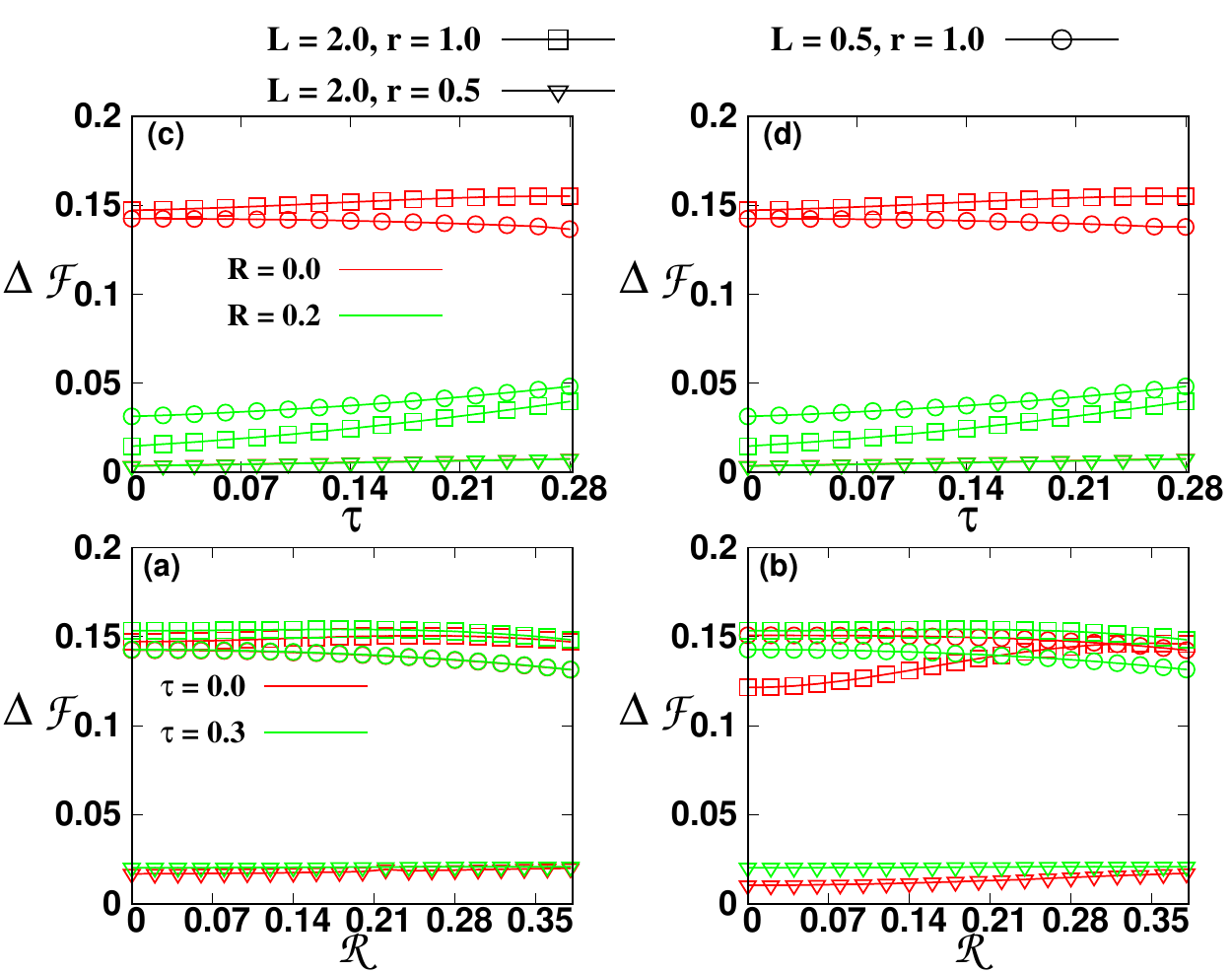}
    \caption{Fidelity deviation, \(\Delta \mathcal{F}\) (vertical axis) against noise parameters, \(\mathcal{R}\) (Bottom),  and \(\tau\) (Top) (horizontal axis) for  squeezed-coherent input states using TMSV states (left panel) and PS states (right panel) as resources. The energy cut-off and resource squeezing are depicted as $L = 2.0, r = 1.0$ (open squares), $L = 2.0, r = 0.5$ (open triangles) and $L = 0.5, r = 1.0$ (open circles).  All other specifications are the same as in Fig. \ref{fig:fig9}. All the axes are dimensionless. }
    \label{fig:fig8}
\end{figure}

 We now shift our attention to the behavior of fidelity deviation against  two noise parameters, \(\mathcal{R}\),  and \(\tau\). In particular, we illustrate the behavior of \(\Delta \mathcal{F}\) with respect to resource noise (measurement noise) at a fixed measurement noise (resource noise) for coherent, squeezed and squeezed coherent states respectively as input.
 
 {\bf Constrained uniform input distribution.} Let us enumerate some of the interesting observations below as depicted in Fig. \ref{fig:fig8}. 
 
 \begin{enumerate} 
 \item {\it Constancy of fidelity deviation.} The first interesting observation is that  $\Delta \mathcal{F}$ remains almost constant with the increase of noise, especially when the variation of \(\mathcal{R}\) for a fixed value of \(\tau\) is studied. A slight increase is seen with the change of \(\tau\). 
 
\item {\it Dependence of input energy on deviation.} $\Delta \mathcal{F}$ possesses a high value for all resource states, across moderate values of $\mathcal{R}$  and $\tau$ when the input energy is high  and the squeezing is moderate.
%It also increases with an increase in the noise parameters. 
%This is expected, since the teleportation protocol worsens with noise and is not very effective at low resource squeezing strengths. 

\item {\it Measurement noise vs. resource noise.} Focusing on the variation of $\Delta \mathcal{F}$ against the measurement inefficiency, we find that at moderate values of the resource noise, e.g. $\tau = 0.3$, the deviation is higher than that of the case with $\tau = 0$ across the entire range of $\mathcal{R}$ except some situations with squeezed coherent input states. Thus there is no counteracting effect of one noise on the other, as in the case of average fidelity. In contrast, if we consider $\Delta \mathcal{F}$ with a nonvanishing moderate value of $\mathcal{R}$, say $0.2$, it is less for all values of $\tau$ for Gaussian as well as non-Gaussian states compared to the situation with vanishing \(\mathcal{R}\) (as depicted in Fig. \ref{fig:fig8}), thereby exhibiting constructive effects also in fidelity deviation. 

%\textcolor{red}{eta constructive to?} 
%\textcolor{red}{This again elucidates the advantage of using non-Gaussianity. Gaussian-eo to dekha jachhe.? } These are generic behaviors exhibited by the fidelity deviation for all input-resource pairs.
 
 %In the following discussion, we will tackle each input ensemble separately, and describe the dynamics of fidelity deviation.
 \item {\it Optimal channels. } Scrutinizing the behavior of fidelity deviation, we observe that even in presence of the noise models considered here, non-Gaussian states give low fidelity deviation compared to that obtained from the Gaussian ones. Among non-Gaussian states, photon added states give low fidelity deviation than that of the photon subtracted ones like the noiseless situation. However, it is important to note that PA states rarely give any quantum advantage according to the average fidelity and hence such a low fidelity deviation does not lead to any benefit in QT. 
 %. However, such an ordering holds for small squeezing strength when the input is the squeezed states while  it always hold for coherent as well as squeezed coherent inputs.  
 
 \end{enumerate}

 \textbf{Role of Gaussian input distribution.} Let us compare the trends of \(\Delta \mathcal{F}\) obtained for inputs chosen from the Gaussian distribution  with the uniform distribution discussed above by varying \(\mathcal{R}\) and \(\tau\). \\
 
 First of all,  the variation of\(\Delta \mathcal{F}\)  observed with \(\mathcal{R}\) and \(\tau\) having low input energy is more than that obtained in constrained uniform case for different inputs.
 %squeezed coherent and squeezed states %while the increasing behavior is observed for coherent input states with \(\tau\). \\
 
\begin{figure}
    \centering
    \includegraphics[width = \linewidth]{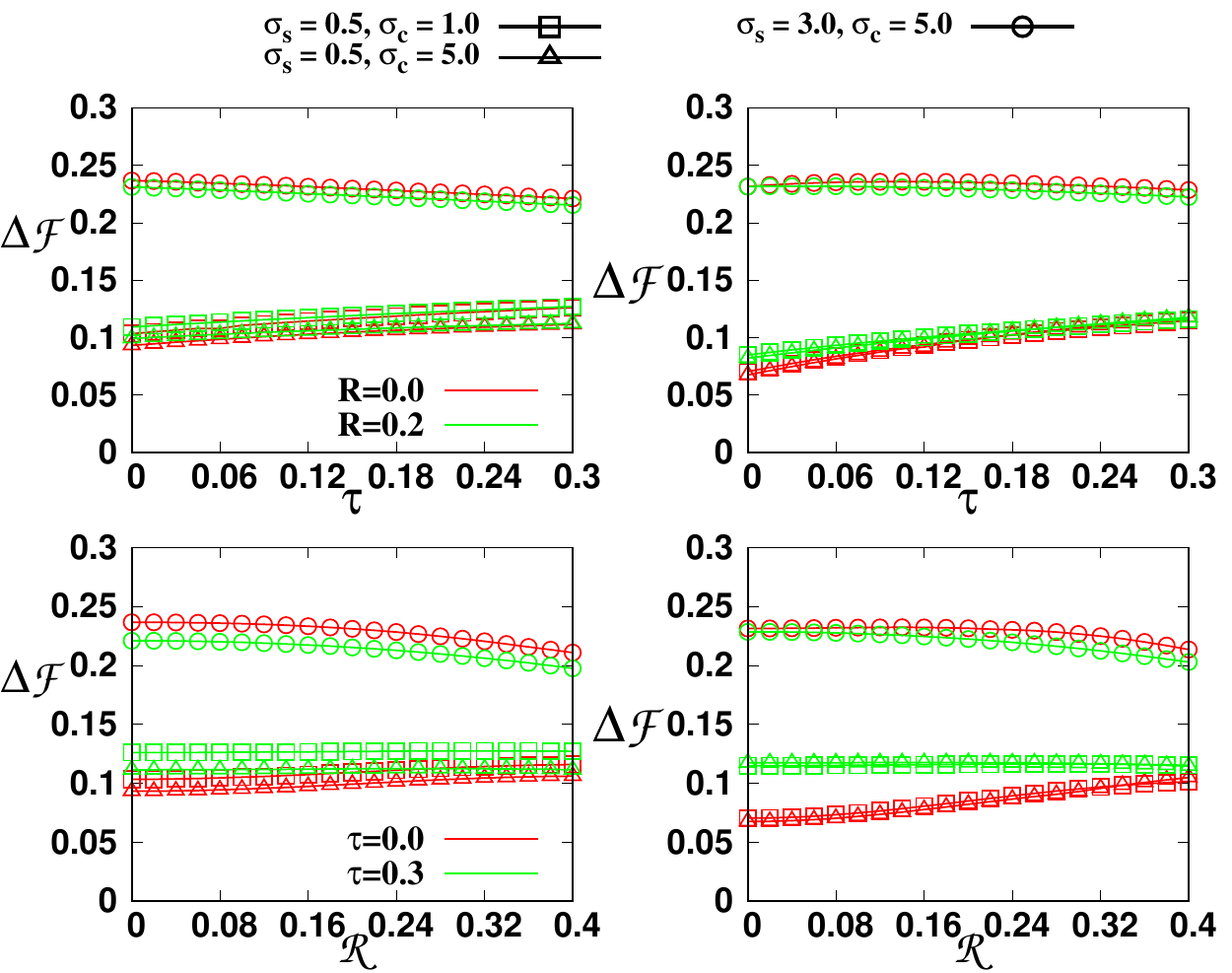}
    \caption{\(\Delta \mathcal{F}\) (ordinate) by varying \(\mathcal{R}\) (Bottom) and \(\tau\) (Top) (abscissa) for  squeezed-coherent states as inputs drawn from a Gaussian distribution using TMSV (left panel) and PS (right panel) as shared channels. We depict the variance parameters as $\sigma_s = 0.5, \sigma_c = 1.0$ (open squares), $\sigma_s = 0.5, \sigma_c = 5.0$ (open triangles), $\sigma_s = 3.0, \sigma_c = 5.0$ (open circles). Here we consider the resource squeezing $r=1.0$. All other specifications are the same as in Fig. \ref{fig:fig9}. All axes are dimensionless.}
    \label{fig:fig14}
\end{figure}

 Secondly, except TMSV states in which high energy sometimes gives low \(\Delta \mathcal{F}\), the relation between input energy and the deviation observed in the uniform distribution remains same  for the Gaussian distribution. \\
 Thirdly, like the uniform distribution, with the increase of \(\tau\) from a vanishing value to a moderate one, deviation always increases for coherent input states while for squeezed coherent state, there are some exceptional regions where the opposite picture emerges  for  all three quantum channels.  However,  unlike uniform distribution, the increase of \(\mathcal{R}\) does not lead to low fidelity deviation in this case with the variation of \(\tau\) -- it  remains almost constant with \(\tau\) after the increase of \(\mathcal{R}\) which can be justified by inspecting Fig. \ref{fig:fig14}. \\

 Finally, analyzing both the fidelity and its deviation along with input energy distributions, one cannot identify a single channel which are more suitable for QT than the others. Specifically, our study reveals that  in presence of noise in measurements as well as  channels,  there is a competition between non-Gaussian photon subtracted  and the Gaussian TMSV states which give the quantum advantage in QT depending on the energy of the input ensembles.

  %\subsubsection{Gaussian distribution}
 %\label{subsubsec:fid_dev_noisy_gau}

  %In upper (lower) panel of Figs. \ref{fig:fig12} and \ref{fig:fig14}, we depict the behaviour of fidelity deviation with respect to resource noise (measurement noise) at a fixed measurement noise (resource noise) for coherent and squeezed coherent states respectively as input. We notice that input squeezed states share more or less similar properties to that of input squeezed-coherent states. In each figure we study our observations for different types of resources- TMSV, PA and PS in the left, middle and right panel respectively.

 % \begin{figure*}
 %    \centering
  %  \includegraphics[width = \linewidth]{fiddev_noisy_coh_gau.pdf}
 %   \caption{for coherent input, TMSV(left),PA(middle),PS(right) }
%    \label{fig:fig12}
%\end{figure*}
% \begin{figure*}
%    \centering
%    \includegraphics[width = \linewidth]{fiddev_noisy_sqz_gau.pdf}
%    \caption{for squeezed input, TMSV(left),PA(middle),PS(right)}
%    \label{fig:fig13}
%\end{figure*}

	\section{Conclusion} 
	\label{sec:conclu}
	
Quantum teleportation is one of the most researched information theoretic protocols, both in terms of its theoretical foundations, as well as experimental implementations. 
Traditionally, the performance of quantum teleportation is assessed using the average fidelity. 
Recently in discrete variable quantum teleportation, it was shown that the standard deviation of fidelity, namely the fidelity deviation, can non-trivially alter the calibration of the performance in teleportation.

In this work, we have introduced the concept of fidelity deviation in continuous variable (CV) quantum teleportation (QT) both for the ideal and noisy cases.
In CV teleportation,  the concept of average fidelity and fidelity deviation, when considered as a direct continuation from the case of discrete variables, suffer from energetic divergences. We  presented regularized versions of these quantities, free from such divergences, by considering that the set of states to be teleported are constrained to have a finite energy cut-off or by introducing Gaussian suppression of the input energy. In particular, for the constrained uniform distribution with a fixed energy threshold,  states are drawn with equal probability over all energy values under the threshold, and for the Gaussian ensemble  a fixed variance determines the average energy range of the input set.

In ideal CV teleportation, we first reported the general trends of average fidelity and fidelity deviation for both the considered constrained uniform and Gaussian distributions of inputs for both Gaussian and non-Gaussian shared states between the sender and the receiver.
In the noiseless scenario, we observed that the average fidelity decreases with the energy of the input state at a fixed value of the resource squeezing. The fidelity deviation too suffers from the rise in ensemble energy, such that it is more for input states of higher energy compared to the inputs having low energy cut-offs. However, the effect of ensemble energy is different on different resource states. We found that the photon added (PA) state is the least useful resource since it can overcome the classical bound only at large values of the input energy. The situation improves for increased resource squeezing, but the photon subtracted (PS) state as well as the  Gaussian TMSV state perform far better than the PA state. The PS state is the most efficient resource since it provides the highest average fidelity for highly energetic input sets
%. It also manages to teleport states 
with reasonably low fidelity deviation, although the PA state furnishes the minimum value in this regard. Overall, advantage is offered by non-Gaussian states for both the figures of merit and the PS state establishes itself as the go-to resource.
%\textcolor{blue}{Then 
We also showed how fidelity deviation can non-trivially alter the hierarchy  among resource states  for which  the average fidelities behave almost identically.

%\sout{The continuous variable paradigm has been largely successful in studying the intricacies of the process and is also more viable in real life experiments.} 

Noise is inevitable  in any experiment, and many developments have been made to study the effect of noise on the primary figure of merit - the average fidelity. We further the investigation into the noisy teleportation process by including the second moment of the fidelity statistics. Our work focuses on the behavior of the aforementioned figures of merit with respect to the input ensembles which are characterized by their energy distribution. We also considered the impact of noise present in the channels as well as measurements on fidelity statistics.
%\sout{The dynamics of teleportation in terms of the input energy is studied by considering two kinds of distributions - the uniform distribution with a fixed cut-off energy, from which states are drawn with equal probability over all energy values under consideration, and the Gaussian ensemble with a fixed standard deviation which determines the average energy of the input set.}
%In the noiseless scenario, we report that the average fidelity decreases with the energy of the input state at a fixed value of the resource squeezing. The fidelity deviation too suffers from the rise in ensemble energy, such that it is more for input states of higher energy. However, the effect of ensemble energy is different on different resource states. We find that the photon added state is the least useful resource since it can overcome the classical bound only at small values of the input energy. The situation improves for increased resource squeezing, but the photon subtracted state as well as the  Gaussian TMSV state perform far better than the PA state. The PS state is the most efficient resource since it provides the highest average fidelity for highly energetic input sets. It also manages to teleport states with reasonably low fidelity deviation, although the PA state furnishes the minimum value in this regard. Overall, advantage is offered by non-Gaussian states for both the figures of merit and the PS state establishes itself as the go-to resource for noiseless CV quantum teleportation.\\
%We showed that the  changes drastically in the presence of noise.
Interestingly, we found that both kinds of noise are seen to affect the non-Gaussian states to a greater extent, in a sense that their average fidelity falls at a much faster rate, thereby making the TMSV state the best resource, especially at higher input energies. The difference in the sources of noise leads to a constructive effect - the resource noise is able to counter the effects of imperfection in measurements, due to which the average fidelity for a higher value of the resource noise is better than that at a lower value of the same, when studied against the variation of the noise. The resource noise also affects the teleportation protocol to a lesser extent, since the figures of merit change very slowly with respect to its variations. 
%The fidelity deviation, however, is the least for the non-Gaussian PA state, although its average fidelity is lower than the classical limit for a large range of the noise parameters. 
Moreover, the effects of noise are less pronounced in case of low energy input ensembles and high squeezing strength of the available resources. In case of the input states, we report that the coherent state suffers much less due to noise, as compared to the squeezed and squeezed-coherent ensembles.

Our work analyzes the performance of CV teleportation protocol in the new light of the regularized version of both average fidelity and fidelity deviation. We demonstrate how incorporating this additional quantifier (fidelity deviation) can provide fundamental insights into the classification of  shared channels for QT that the average fidelity alone cannot capture  both in noiseless and noisy scenarios. We believe that the present work opens new avenues into research on CV teleportation.

%\textcolor{red}{The efficacy of the resources have previously not been studied in terms of the input energy}
%\textcolor{blue}{journals typically don't allow statements of newness, i.e., amra age korechi....so better avoid that!}

%and the present work opens new avenues into the research on CV teleportation based on the input states to be transferred.

\acknowledgements

We acknowledge the support from the Interdisciplinary Cyber Physical Systems (ICPS) program of the Department of Science and Technology (DST), India, Grant No.: DST/ICPS/QuST/Theme- 1/2019/23, the use of \href{https://github.com/titaschanda/QIClib}{QIClib} -- a modern C++ library for general purpose quantum information processing and quantum computing (\url{https://titaschanda.github.io/QIClib}).

	\bibliography{bib}
	
\end{document}